\documentclass[twoside,twocolumn,9pt]{article}
\usepackage{extsizes}
\usepackage[super,sort&compress,comma]{natbib}
\usepackage[version=3]{mhchem}
\usepackage[left=1.5cm, right=1.5cm, top=1.785cm, bottom=2.0cm]{geometry}
\usepackage{balance}
\usepackage{mathptmx}
\usepackage{sectsty}
\usepackage{graphicx}
\usepackage{multirow}
\usepackage[table,xcdraw]{xcolor}
\usepackage{lastpage}
\usepackage{amsmath}
\usepackage{amssymb}
\usepackage[format=plain,justification=justified,singlelinecheck=false,font={stretch=1.125,small,sf},labelfont=bf,labelsep=space]{caption}
\usepackage{float}
\usepackage[hidelinks,draft]{hyperref}
\usepackage{fancyhdr}
\usepackage{fnpos}
\usepackage{cleveref}
\usepackage[english]{babel}
\addto{\captionsenglish}{%
  
}
\usepackage{array}
\usepackage{droidsans}
\usepackage{charter}
\usepackage{setspace}
\usepackage[compact]{titlesec}
\usepackage{xspace}
\usepackage{subcaption}
\usepackage{braket}
\usepackage{multirow}
\usepackage{bm}
\usepackage[floats=float]{chemscheme}
\crefname{figure}{Figure}{Figures}
\crefname{table}{Table}{Tables}
\crefname{equation}{Eq.}{Eqs.}
\crefname{section}{Section}{Sections}
\crefname{subsection}{Section}{Sections}

\usepackage{epstopdf}%This line makes .eps figures into .pdf - please comment out if not required.

\definecolor{cream}{RGB}{222,217,201}

\begin{document}

\pagestyle{fancy}
\thispagestyle{plain}
\fancypagestyle{plain}{
%%%HEADER%%%
\renewcommand{\headrulewidth}{0pt}
}
%%%END OF HEADER%%%

%%%PAGE SETUP - Please do not change any commands within this section%%%
\makeFNbottom
\makeatletter
\renewcommand\LARGE{\@setfontsize\LARGE{15pt}{17}}
\renewcommand\Large{\@setfontsize\Large{12pt}{14}}
\renewcommand\large{\@setfontsize\large{10pt}{12}}
\renewcommand\footnotesize{\@setfontsize\footnotze{7pt}{10}}
\makeatother

\renewcommand{\thefootnote}{\fnsymbol{footnote}}
\renewcommand\footnoterule{\vspace*{1pt}%
\color{cream}\hrule width 3.5in height 0.4pt \color{black}\vspace*{5pt}}
\setcounter{secnumdepth}{5}

\makeatletter
\renewcommand\@biblabel[1]{#1}
\renewcommand\@makefntext[1]%
{\noindent\makebox[0pt][r]{\@thefnmark\,}#1}
\makeatother
\renewcommand{\figurename}{\small{Fig.}~}
\sectionfont{\sffamily\Large}
\subsectionfont{\normalsize}
\subsubsectionfont{\bf}
\setstretch{1.125} %In particular, please do not alter this line.
\setlength{\skip\footins}{0.8cm}
\setlength{\footnotesep}{0.25cm}
\setlength{\jot}{10pt}
\titlespacing*{\section}{0pt}{4pt}{4pt}
\titlespacing*{\subsection}{0pt}{15pt}{1pt}
%%%END OF PAGE SETUP%%%

%%%FOOTER%%%
\fancyfoot{}
\fancyfoot[LO,RE]{\vspace{-7.1pt}\includegraphics[height=9pt]{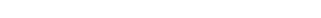}}
\fancyfoot[CO]{\vspace{-7.1pt}\hspace{11.9cm}\includegraphics{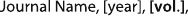}}
\fancyfoot[CE]{\vspace{-7.2pt}\hspace{-13.2cm}\includegraphics{head_foot/RF}}
\fancyfoot[RO]{\footnotesize{\sffamily{1--\pageref{LastPage} ~\textbar  \hspace{2pt}\thepage}}}
\fancyfoot[LE]{\footnotesize{\sffamily{\thepage~\textbar\hspace{4.65cm} 1--\pageref{LastPage}}}}
\fancyhead{}
\renewcommand{\headrulewidth}{0pt}
\renewcommand{\footrulewidth}{0pt}
\setlength{\arrayrulewidth}{1pt}
\setlength{\columnsep}{6.5mm}
\setlength\bibsep{1pt}
%%%END OF FOOTER%%%

%%%FIGURE SETUP - please do not change any commands within this section%%%
\makeatletter
\newlength{\figrulesep}
\setlength{\figrulesep}{0.5\textfloatsep}

\newcommand{\topfigrule}{\vspace*{-1pt}%
\noindent{\color{cream}\rule[-\figrulesep]{\columnwidth}{1.5pt}} }

\newcommand{\botfigrule}{\vspace*{-2pt}%
\noindent{\color{cream}\rule[\figrulesep]{\columnwidth}{1.5pt}} }

\newcommand{\dblfigrule}{\vspace*{-1pt}%
\noindent{\color{cream}\rule[-\figrulesep]{\textwidth}{1.5pt}} }

\makeatother
%%%END OF FIGURE SETUP%%%

%%%TITLE, AUTHORS AND ABSTRACT%%%
\twocolumn[
  \begin{@twocolumnfalse}
{\includegraphics[height=30pt]{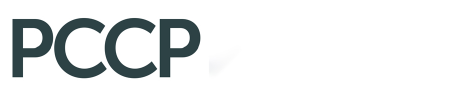}\hfill\raisebox{0pt}[0pt][0pt]{\includegraphics[height=55pt]{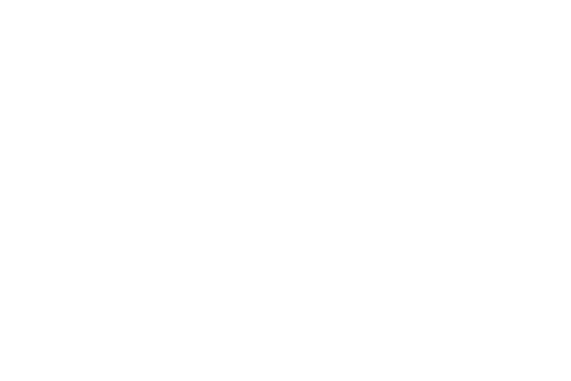}}\\[1ex]
\includegraphics[width=18.5cm]{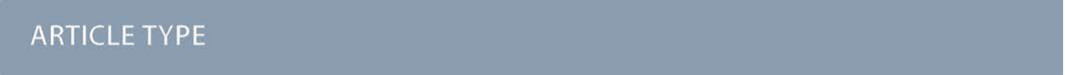}}\par
\vspace{1em}
\sffamily
\begin{tabular}{m{4.5cm} p{13.5cm} }

\includegraphics{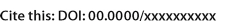} & \noindent\LARGE{\textbf{Decoding Transient X-ray Absorption Spectra of \mbox{Acetylacetone} With Multireference Algebraic Diagrammatic Construction Theory
}} \\
\vspace{0.3cm} & \vspace{0.3cm} \\

 & \noindent\large{Bennett~W.~Clark,\textit{$^{a}$} Donna~H.~Odhiambo,\textit{$^{a}$} Haden~Dickerson,\textit{$^{a}$} and Alexander Yu.~Sokolov$^{\ast}$\textit{$^{a}$}} \\

\includegraphics{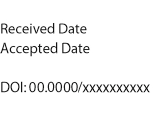} & \noindent\normalsize{
Time-resolved X-ray absorption spectroscopy (TR-XAS) offers an element- and site-specific probe of coupled electronic and nuclear dynamics, but its interpretation requires methods that can treat multiconfigurational excited states across nonequilibrium nuclear ensembles.
Here, we report time-resolved simulations of the acetylacetone (AcAc) transient X-ray absorption spectra by combining surface-hopping dynamics from complete active space second-order perturbation theory with multireference algebraic diagrammatic construction (MR-ADC).
The simulated 20--200 fs spectra show a good agreement with experimental measurements and reveal that the TR-XAS response is governed by continuously evolving distributions of molecular geometries on the singlet potential energy surfaces ($S_2$ and $S_1$).
In particular, absorption at 279.5--281.5 eV is enhanced for transient, nearly symmetric proton-sharing configurations, whereas the 284--286 eV profile reflects geometry-dependent C~1s excitations modulated by proton transfer, bond alternation, and ring opening.
For the long-time $T_1$ spectrum (7--10 ps), the principal features near 281.4 and 283.8 eV are assigned to central C 1s excitations into low-lying triplet $\pi$ orbitals.
Together, our simulations provide a more complete mechanistic picture of AcAc photorelaxation by directly linking ultrafast carbon K-edge signals to proton transfer, skeletal reorganization, internal conversion, and triplet formation.
} \\

\end{tabular}

 \end{@twocolumnfalse} \vspace{0.6cm}

  ]
%%%END OF TITLE, AUTHORS AND ABSTRACT%%%

%%%FONT SETUP - please do not change any commands within this section
\renewcommand*\rmdefault{bch}\normalfont\upshape
\rmfamily
\section*{}
\vspace{-1cm}

%%%FOOTNOTES%%%

\footnotetext{\textit{$^{a}$~Department of Chemistry and Biochemistry, The Ohio State University, Columbus, Ohio, 43210, USA. E-mail: sokolov.8@osu.edu}}

%Please use \dag to cite the ESI in the main text of the article.
%If you article does not have ESI please remove the \dag symbol from the title and the footnotetext below.
\footnotetext{\dag~Electronic Supplementary Information (ESI) available: spectral benchmarking, active spaces, trajectory sampling analysis, and full TR-XAS spectrum simulated using MR-ADC(2). See DOI: }

%additional addresses can be cited as above using the lower-case letters, c, d, e... If all authors are from the same address, no letter is required

%\footnotetext{\ddag~Additional footnotes to the title and authors can be included \textit{e.g.}\ `Present address:' or `These authors contributed equally to this work' as above using the symbols: \ddag, \textsection, and \P. Please place the appropriate symbol next to the author's name and include a \texttt{\textbackslash footnotetext} entry in the correct place in the list.}

%%%END OF FOOTNOTES%%%

\raggedbottom

\section{Introduction}

Time-resolved X-ray absorption spectroscopy (TR-XAS) is an ultrafast pump--probe technique that provides element-specific information about electronic structure and photo-induced relaxation dynamics.
In a typical TR-XAS experiment, an ultrafast UV pump pulse excites the system from the ground electronic state,\cite{stolow:2004p1719} and a delayed X-ray probe promotes electrons from core orbitals into unoccupied valence orbitals, Rydberg-like states, or the ionization continuum.
Because core-level absorption energies are element specific, TR-XAS can track changes in local electronic and geometric structure following photoexcitation.\cite{loh:2013p292}
Recent advances in femtosecond and attosecond X-ray sources\cite{hentschel:2001p509, loh:2013p292} have made it possible to follow coupled electronic and nuclear motion in real time, enabling applications ranging from catalysis\cite{lin:2020p3525} and battery materials\cite{miao:2025pe01608} to reaction dynamics\cite{kim:2016p3734} and excited-state relaxation pathways.\cite{bhattacherjee:2017p16576,haugen:2024p,haugen:2023p634,severino:2025p30785}
These experiments aim to provide a ``molecular movie'' of how electronic and nuclear structure evolve after vertical excitation.\cite{kraus:2018p82}

%figure - AcAc Geometries x
\begin{figure}[t!]
	\centering
	\includegraphics[width=0.8\columnwidth]{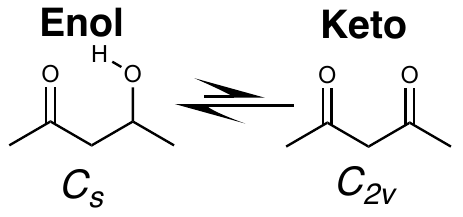}
	\caption{In the gas phase, tautomeric equilibrium for acetylacetone favors the enol tautomer with $C_{s}$ symmetry.}
	\label{fig:enol_keto_acac_hfacac}
\end{figure}
%end figure

\begin{figure}[t!]
	\centering
	\includegraphics[width=1.0\columnwidth]{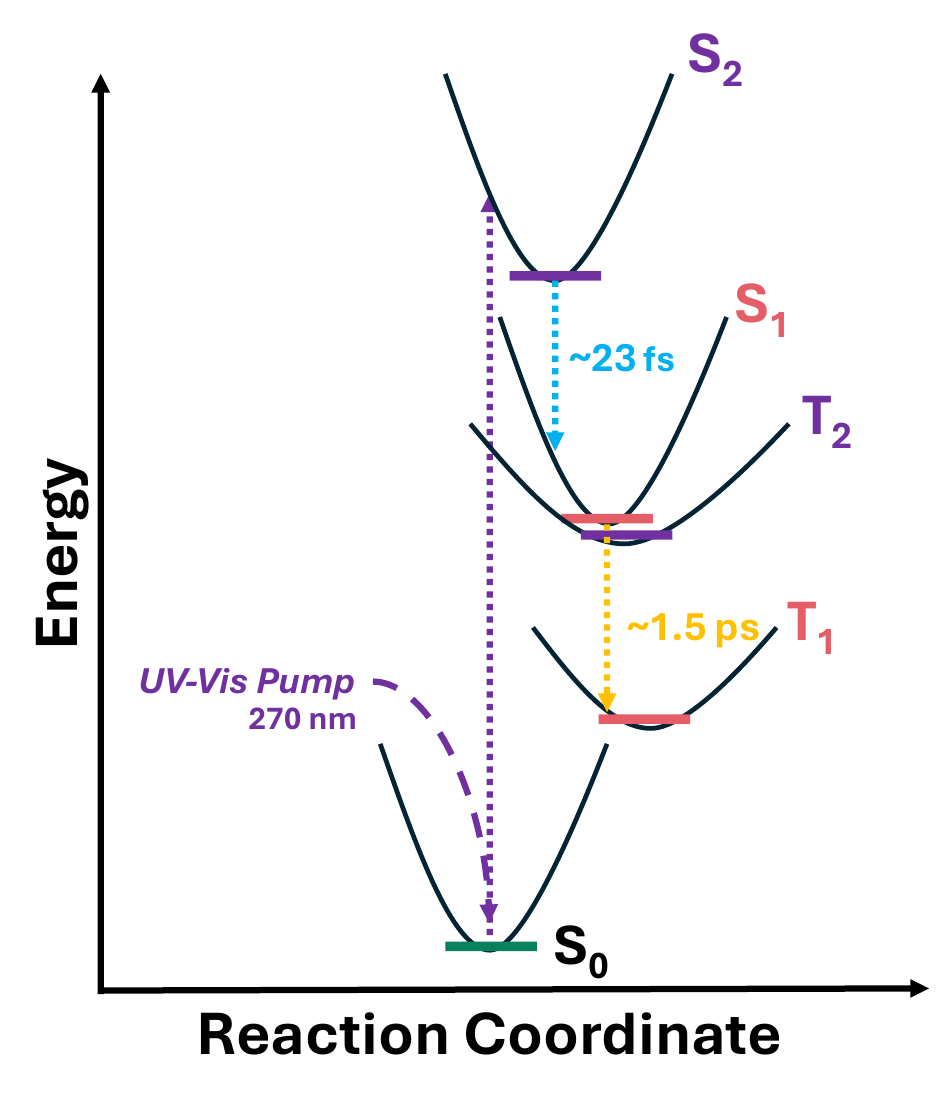}
	\caption{Schematic illustration\cite{squibb:2018p63} of the gas-phase photorelaxation pathway of AcAc, as inferred from recent time-resolved spectroscopic studies.\cite{bhattacherjee:2017p16576,severino:2025p30785} 
	Excitation with 270 nm light initially promotes AcAc from the ground state, $S_0$, to the $\pi\pi^\ast$-type $S_2$ state (purple). 
	The molecule then undergoes rapid internal conversion from $S_2$ to the $n\pi^\ast$ $S_1$ state through excited-state intramolecular hydrogen transfer (ESIHT). 
	Subsequent twisting facilitates intersystem crossing (yellow) from $S_1$ to the $T_1$ state on a 1.5 ps timescale.
	}
	\label{fig:photodissociation_diagram}
\end{figure}
%end figure

Acetylacetone (AcAc) has recently emerged as a prototypical system for TR-XAS studies because it exhibits unusually rich excited-state dynamics for a small organic molecule.\cite{bhattacherjee:2017p16576,haugen:2024p, severino:2025p30785, scutelnic:2025p5068}
In the gas phase, AcAc exists predominantly in its enol tautomeric form (\cref{fig:enol_keto_acac_hfacac}), which is stabilized by a strong intramolecular hydrogen bond.\cite{nakanishi:1977p2255,lowrey:1971p6399,andreassen:1972p381,folkendt:1985p3347,folkendt:1985p3347,dolati:2016p344}
Although the equilibrium geometry of the gas-phase enol has been debated, most recent theoretical studies identify an asymmetric $C_s$ structure as the global energy minimum.\cite{dolati:2016p344,chen:2006p4434,caminati:2006p854,severino:2025p30785,squibb:2018p63,belova:2014p5412}

Following excitation at 270 nm, AcAc is promoted from the ground $S_0$ state to the bright $\pi\pi^\ast$ $S_2$ state. 
It subsequently undergoes ultrafast internal conversion to the $n\pi^\ast$ $S_1$ state within $23 \pm 2$ fs, followed by intersystem crossing to the $T_1$ state on a timescale of approximately 1.5 ps (\cref{fig:photodissociation_diagram}).\cite{bhattacherjee:2017p16576,severino:2025p30785} 
The initial internal-conversion dynamics are driven by excited-state intramolecular hydrogen transfer (ESIHT), whereas the subsequent intersystem crossing has been linked to an out-of-plane twisting motion.\cite{severino:2025p30785}
This multistep relaxation pathway has been investigated using complementary ultrafast techniques, including TR-XAS,\cite{bhattacherjee:2017p16576,haugen:2024p} ultraviolet and extreme-ultraviolet time-resolved photoelectron spectroscopy,\cite{kotsina:2020p4647,severino:2025p30785,pini:2026p9912} and ultrafast electron diffraction,\cite{scutelnic:2025p5068} together with high-level electronic structure calculations.\cite{chen:2006p4434,squibb:2018p63,severino:2025p30785, chakraborty:2026p1090}
Despite this progress, a reliable microscopic assignment of the transient spectral features remains challenging because the observed spectra simultaneously encode changes in electronic-state character, core-hole relaxation and correlation, nonadiabatic nuclear motion, and vibrational averaging.

The difficulties encountered for AcAc exemplify the broader theoretical challenges of simulating TR-XAS, which are substantially greater than those associated with ground-state XAS. 
Whereas a ground-state spectrum can often be modeled at a single equilibrium geometry, TR-XAS simulations must account for evolving excited-state populations and nuclear configurations while accurately describing core-level excitations, core-hole screening, double-excitation character, relativistic effects, and both static and dynamic electron correlation.\cite{schuurman:2022p20012} 
Common single-reference approaches, including time-dependent density functional theory,\cite{gutierrez-quintanilla:2023p158} equation-of-motion coupled-cluster theory,\cite{faber:2019p144107,tsuru:2021p024101} and many-body perturbation theory,\cite{forster:2022p6779} have been widely applied to transient spectra.
However, these methods can become unreliable when the relevant valence or core-excited states possess strong multiconfigurational character.
Multireference approaches such as complete active space self-consistent field methods,\cite{helmich-paris:2021pe26559} CASPT2,\cite{andersson:1992p1218} and NEVPT2\cite{angeli:2002p9138, freitag:2017p451} can account for static correlation, but their conventional formulations are generally limited to excitations within a selected set of frontier orbitals, known as the active space. 
Restricted-active-space approaches such as RASSCF and RASPT2 can incorporate core-hole configurations more directly,\cite{montorsi:2022p1003,tran:2019p5223} but suffer from intruder-state problems, convergence difficulties, and artifacts of state-averaging.
These challenges have spurred substantial efforts to develop multireference methods for accurately simulating X-ray absorption spectra.\cite{seidu:2019p144104, maganas:2019p104106, huang:2025p6834}

In this work, we employ multireference algebraic diagrammatic construction (MR-ADC) theory to simulate and interpret the transient X-ray absorption spectra of AcAc.\cite{sokolov:2018p204113,chatterjee:2019p5908,chatterjee:2020p6343,mazin:2021p6152,banerjee:2023p3037}
MR-ADC combines an explicit treatment of excitations involving all (core, valence, and virtual) molecular orbitals with accurate description of static and dynamic correlation while avoiding intruder-state and state-averaging problems.
When combined with the core--valence separation (CVS) approximation,\cite{cederbaum:1980p206,barth:1981p1038} it efficiently targets core-excited and core-ionized states.\cite{demoura:2022p8041,demoura:2022p4769,mazin:2023p4991,demoura:2024p5816}
Previous CVS-MR-ADC applications have shown good agreement with experiment and accurate theoretical benchmarks. 
Representative examples include time-resolved X-ray photoelectron spectra (XPS) of Fe(CO)$_5$ and its photodissociation products,\cite{gaba:2024p15927} the XPS and XAS spectra of ozone,\cite{mazin:2023p4991} and the XPS spectra of substituted ferrocenes.\cite{demoura:2024p5816} 

Here, we benchmark MR-ADC predictions for AcAc against available experimental data,\cite{bhattacherjee:2017p16576,haugen:2024p,severino:2025p30785}  compare our results to previous theoretical calculations,\cite{bhattacherjee:2017p16576,faber:2019p144107,tsuru:2021p024101,arias-martinez:2024p752} and use the method to assign transient XAS features along the photorelaxation pathway.
To account for nuclear motion, we propagate nonadiabatic molecular dynamics trajectories, sample nonequilibrium geometries, compute XAS spectra for the sampled structures, and average the resulting intensities.
Comparison with spectra evaluated at representative stationary geometries reveals how nuclear motion modifies transient peak positions, intensities, and line shapes.
To our knowledge, this is the first study of AcAc photorelaxation in which the nonadiabatic dynamics, electronic structure, and transient XAS observables are all treated at multireference levels of theory.
The resulting vibrationally averaged spectra enable assignments of features that were not fully resolved in previous TR-XAS measurements.
Together with recent TR-XAS,\cite{bhattacherjee:2017p16576,haugen:2024p} time-resolved XUV photoelectron,\cite{kotsina:2020p4647,severino:2025p30785,pini:2026p9912} and ultrafast electron-diffraction\cite{scutelnic:2025p5068} studies, our results provide a more complete microscopic picture of the coupled electronic and nuclear dynamics governing AcAc photorelaxation.

%%%%%%%%%%%%%%%%%%%%%%%%%%%%%%%%%%%%%%%%%%%%%%%%%%%%%%%%%%%%%%%%%%%%%
% 2. Theory
%% 2.1. Outline of ADC: ADC theory, Green's Functions, etc
%% 2.2. Outline of MR-ADC: CAS reference, Dyall Hamiltonian, Eigenvalue problem
%% 2.3. MR-ADC flavors: MR-ADC(2) and MR-ADC(2)-X
%% 2.4. CVS-IP-MR-ADC: Core-Valence Separation Approach and previous papers (CVS-IP and CVS-EE)
%%%%%%%%%%%%%%%%%%%%%%%%%%%%%%%%%%%%%%%%%%%%%%%%%%%%%%%%%%%%%%%%%%%%%
\section{Theory}
\label{sec:Theory}

\subsection{Simulating X-ray Absorption Spectra With Multireference Algebraic Diagrammatic Construction Theory}
\label{sec:Theory:mr_adc}

Accurate simulations of TR-XAS spectra require an electronic structure method capable of describing non-equilibrium geometries, nonadiabatic topologies, and dense manifolds of mixed excitations, such as core--valence, core--Rydberg, and doubly excited transitions.
Such a method must also provide a balanced treatment of electron correlation across these excited states. %landscape or topography?
While traditional multireference perturbation theory (MRPT) methods can efficiently capture electron correlation, their restricted orbital excitation range, intruder states susceptibility, and empirical parameter dependence limit their applicability to probing core-excited states along complex photophysical pathways.\cite{sarkar:2022p2418} %cite roos:1995p215 for ISP?
Multireference algebraic diagrammatic construction (MR-ADC) circumvents these limitations by combining the rigorous correlation treatment and efficiency of MRPT with an inherently multistate and intruder-state-free Green's function framework.\cite{sokolov:2018p204113, sokolov:2024p121}
Additionally, MR-ADC is Hermitian, size-consistent, and size-extensive, making it particularly well suited for modeling the excited-state dynamics probed in TR-XAS experiments. Here, we present an overview of MR-ADC for computing core excitation energies and simulating X-ray absorption spectra.
A more rigorous discussion of the theory can be found elsewhere.\cite{banerjee:2023p3037, sokolov:2018p204113, chatterjee:2019p5908, chatterjee:2020p6343, demoura:2022p8041, demoura:2022p4769, sokolov:2024p121}
%\cite{sokolov:2018p204113,chatterjee:2019p5908,banerjee:2019p224112, chatterjee:2020p6343,demoura:2022p4769,sokolov:2024p121} 

To simulate electronic excitations, MR-ADC approximates the frequency-dependent polarization propagator, $\Pi(\omega)$, which describes the linear response of a system in initial state $\ket{\Psi_0}$ with energy $E_0$ to an external perturbation of frequency $\omega$.\cite{fetter:1971p, dickhoff:2008p}
Generally, $\boldsymbol{\Pi}(\omega)$ can be decomposed into the sum of forward and backward components: $\boldsymbol{\Pi} (\omega) = \boldsymbol{\Pi}^{+}(\omega) + \boldsymbol{\Pi}^{-}(\omega)$. Since these components are interrelated and contain the same physical information, it is convention to consider only the forward component, such that:\cite{schirmer:2018p195,trofimov:2006p1,oddershede:1984p33,oddershede:1992p303}
\begin{equation}
	\label{eq:pp}
	\Pi_{\mu\nu} (\omega) \equiv \Pi_{\mu\nu}^+ (\omega) =  \braket{\Psi | q_\mu (\omega - H + E)^{-1} q_\nu^\dag | \Psi}
\end{equation}
where $q_\nu^\dag = a_p^\dag a_q - \braket{\Psi |a_p^\dag a_q | \Psi}$ is the perturbation operator, $a_p^\dag$ and $a_q$ are the fermionic creation and annihilation operators, $\ket{\Psi}$ is the exact ground-state wavefunction, and $E$ is the exact ground-state energy of Hamiltonian $H$.
Importantly, the poles of $\boldsymbol{\Pi}(\omega)$ correspond to the exact excitation energies ($\omega_n = E_n - E_0$), while the associated residues correspond to the transition probabilities ($ X_{\mu n} X_{\nu n}^\dagger = \braket{\Psi_0 | q_\mu |\Psi_n}\braket{\Psi_n | q_\nu^\dagger | \Psi_0}$).

In practice, the exact eigenstate basis is unknown. MR-ADC instead approximates the propagator using a non-orthogonal basis of excited states constructed from the multiconfigurational $N$-electron reference wavefunction obtained from a complete active space self-consistent field (CASSCF) calculation.\cite{hinze:1973p6424, siegbahn:1980p1229, roos:1980p157, siegbahn:1981p2384, werner:1981p5794, werner:1985p5053} 
In compact matrix form, this approximation can be expressed as a product of nondiagonal matrices:
\begin{equation}
	\boldsymbol{\Pi}(\omega) = \mathbf{T} (\omega \mathbf{S} - \mathbf{M})^{-1} \mathbf{T^\dagger},
\end{equation}
Here, the effective Hamiltonian matrix $\mathbf{M}$ and effective transition moments matrix $\mathbf{T}$ encode the poles and residues of \cref{eq:pp}, respectively, while the overlap matrix $\mathbf{S}$ accounts for the nonorthogonality of the basis.
The MR-ADC($n$) hierarchy is defined by the independent perturbative expansion of $\mathbf{M}$, $\mathbf{T}$, and $\mathbf{S}$ through $n$th order in perturbation theory.
The excitation energies $\boldsymbol{\Omega}$ and eigenvectors $\mathbf{Y}$ are then obtained by solving the generalized eigenvalue problem $\mathbf{M Y} = \mathbf{S Y}\boldsymbol{\Omega}$,
from which the approximate spectroscopic amplitudes are computed as $\mathbf{X} = \mathbf{TS}^{-1/2}\mathbf{Y}$.

Together, $\boldsymbol{\Omega}$ and $\mathbf{X}$ define the spectral representation of the MR-ADC($n$) propagator, from which the spectral function is obtained as
\begin{equation}
	\label{eq:spec_fcn}
	A(\omega)
	=
	-\frac{1}{\pi}
	\operatorname{Im}
	\left[
	\operatorname{Tr}
	\left(
	\mathbf{D}^{\dagger}
	\boldsymbol{\Pi}(\omega)
	\mathbf{D}
	\right)
	\right].
\end{equation}
Here, $\mathbf{D}$ denotes the matrix representation of dipole-moment operator.
Within the electric-dipole approximation, $A(\omega)$ is directly related to the corresponding absorption spectrum.

To target core excitations within the MR-ADC framework, we employ the core--valence separation (CVS) scheme,\cite{cederbaum:1980p206, barth:1981p1038, cederbaum:1987p622} which exploits the energetic and spatial separation between core and valence electrons to decouple their excitation spaces. This decoupling is achieved by restricting the excitation space to configurations involving at least one core orbital.
\begin{figure}[t!]
	\centering
	\includegraphics[width=0.49\textwidth]{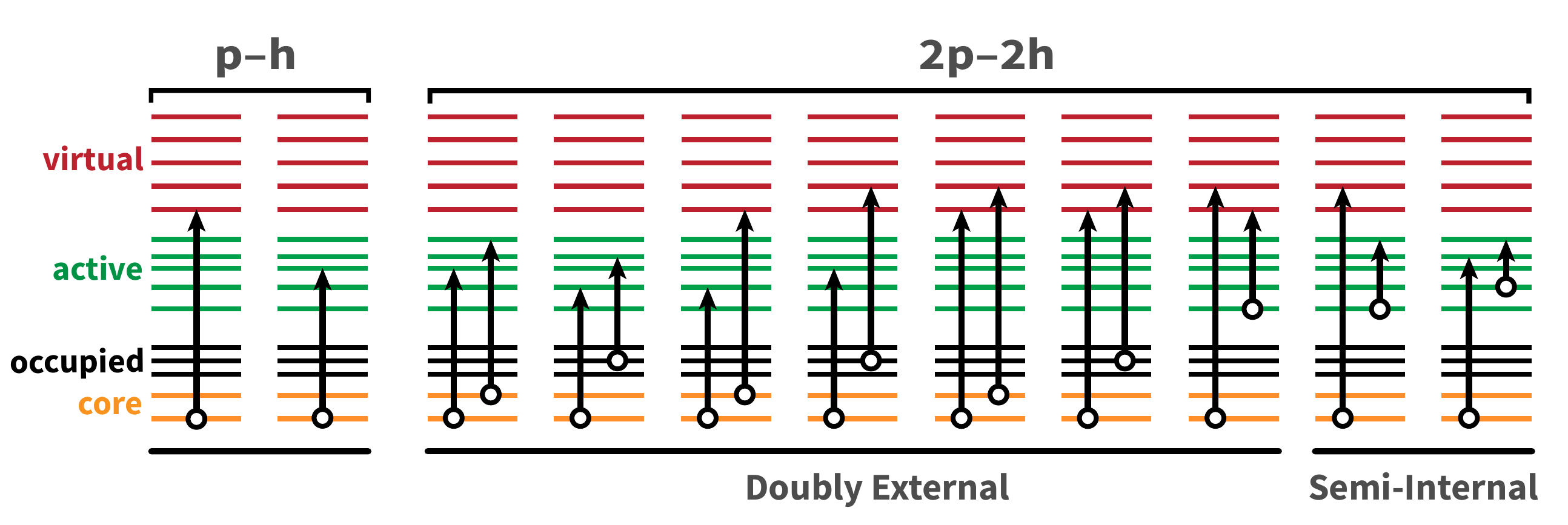}
	\captionsetup{justification=justified,singlelinecheck=false,font=footnotesize}
	\caption{Excited electronic configurations representing matrix elements of the effective Hamiltonian $\mathbf{M}$ (\cref{fig:m_matrices}). Circles with arrows denote single-electron excitations. Yellow, black, green, and red lines represent core, (doubly) occupied, active, and virtual orbitals, respectively.
	%{\color{gray} [Reproduced with permission from Ref.\@\citenum{mazin:2023p4991}]}
	}
	\label{fig:adc_diagram}
\end{figure}

In second-order MR-ADC, this basis comprises particle--hole (p--h) and two-particle--two-hole (2p--2h) configurations (\cref{fig:adc_diagram}). The underlying orbital space derives from the CASSCF reference state, which partitions molecular orbitals into (doubly) occupied, active, and external (unoccupied) subspaces. CVS further subdivides the occupied space into low-lying core orbitals (from which excitations are required) and valence-occupied orbitals.

\begin{figure}[t!]
	\centering
	\includegraphics[width=0.49\textwidth]{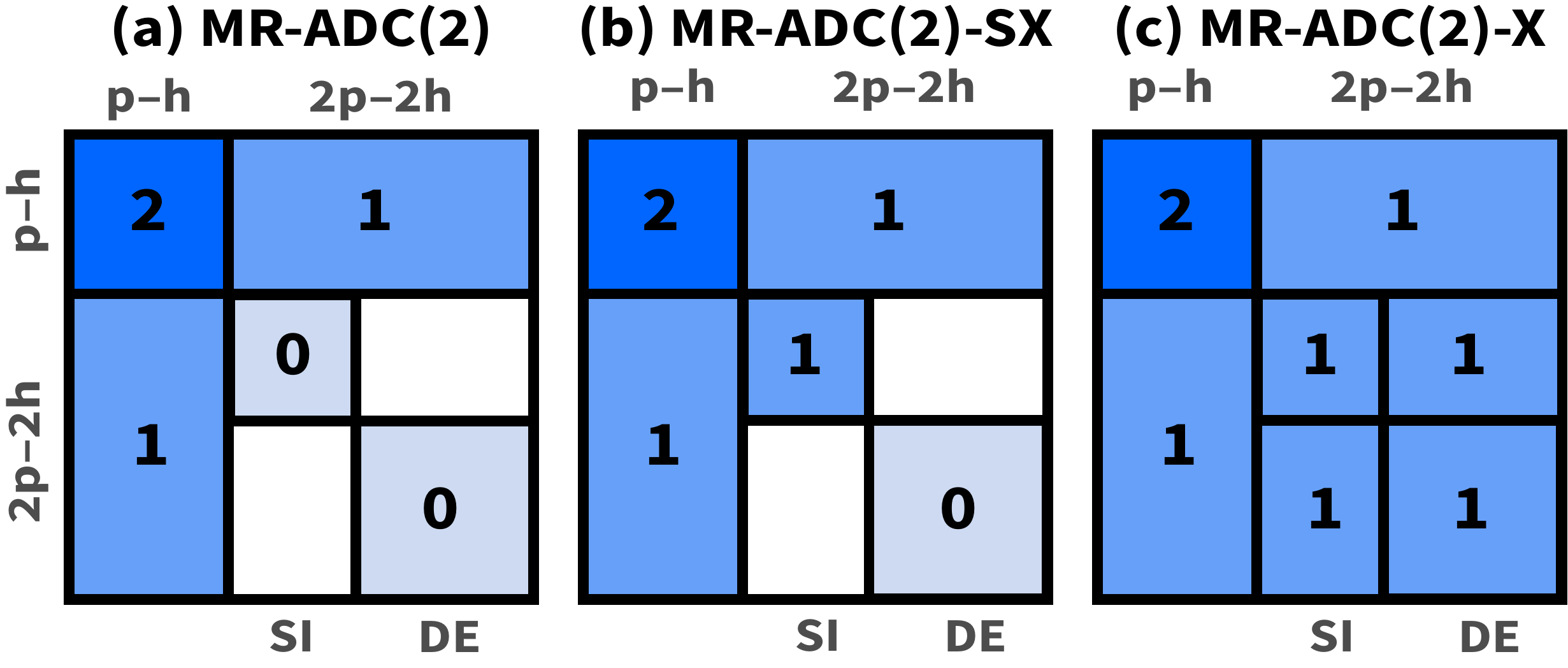}
	\captionsetup{justification=justified,singlelinecheck=false,font=footnotesize}
	\caption{		
		Perturbative structure of the effective Hamiltonian matrix \textbf{M} in (a) MR-ADC(2), (b) MR-ADC(2)-SX, and (c) MR-ADC(2)-X. 
		See \cref{fig:adc_diagram} for schematic representation of all particle--hole (p-h) and two-particle--two-hole (2p-2h) excitations.
		SI and DE denote semi-internal and doubly external 2p-2h excitations.
		The shaded areas denote nonzero blocks.
	}
	\label{fig:m_matrices}
\end{figure}

The second-order MR-ADC framework admits several variants that differ in the perturbative treatment of couplings within the two-particle--two-hole (2p--2h) excitation manifold.
In the strict MR-ADC(2) formulation, these couplings are retained only through zeroth order, whereas the extended second-order variant (MR-ADC(2)-X) extends the treatment to first order for the full 2p--2h sector.
Here, we also introduce an intermediate semi-internal extended approximation, denoted MR-ADC(2)-SX, in which first-order corrections are included selectively for semi-internal 2p--2h configurations, while the remaining 2p--2h couplings are treated at zeroth order.
By incorporating higher-order effects for double excitations, the extended variants provide an improved description of orbital relaxation in particle--hole states and of the energies and intensities of satellite transitions, while preserving the favorable computational scaling of MR-ADC(2).
The perturbative block structure of the effective Hamiltonian, $\mathbf{M}$, for the three approximations is summarized in \cref{fig:m_matrices}.

\subsection{Time-Resolved XAS Simulations From Multireference Surface-Hopping Dynamics}

To generate the nonequilibrium nuclear ensembles required for the TR-XAS simulations, we employed fewest-switches surface hopping (FSSH),\cite{tully:1990p1061,mukherjeePredictionChallengeSimulating2024,jainPedagogicalOverviewFewest2022,mai:2025p,mai:2018pe1370} a mixed quantum--classical approach in which the nuclei evolve classically on adiabatic potential-energy surfaces while the electronic wavefunction is propagated quantum mechanically.
Here, we briefly summarize the aspects of FSSH relevant to the present simulations and its combination with MR-ADC spectroscopy.

For a trajectory evolving on electronic state $k$, the nuclear coordinates $\mathbf{R}$ are propagated according to Newton's equations,
\begin{equation}
	m\ddot{\mathbf{R}}
	=
	-\bra{\psi_k}\nabla_{\mathbf{R}}\hat{H}_{\mathrm{el}}\ket{\psi_k},
\end{equation}
where $\hat{H}_{\mathrm{el}}$ is the electronic Hamiltonian, $\ket{\psi_k}$ is the adiabatic electronic wavefunction, $m$ denotes the nuclear masses, and $\ddot{\mathbf{R}}$ represents the nuclear accelerations.
Transitions between electronic states are driven by the nonadiabatic derivative couplings,
\begin{equation}
	\mathbf{d}_{kj}
	=
	\braket{\psi_k | \nabla_{\mathbf{R}} \psi_j}.
\end{equation}
At each time step, the probability of a hop from the active state $k$ to another state $j$ is evaluated according to the fewest-switches prescription,
\begin{equation}
	g_{kj}
	=
	\max
	\left[
	\frac{2\,\mathrm{Re}
		\left(
		c_k^{*}c_j\,
		\mathbf{v}\cdot\mathbf{d}_{kj}
		\right)}
	{|c_k|^2}
	\Delta t,
	0
	\right],
\end{equation}
where $c_k$ and $c_j$ are the electronic-state amplitudes, $\mathbf{v}$ is the nuclear velocity, and $\Delta t$ is the time step.
The resulting hopping probability is compared with a random number to determine whether the trajectory changes its active electronic state.
Because FSSH is an ensemble method, an ensemble of initial conditions is propagated to recover the time-dependent electronic-state populations and distribution of nuclear geometries.
Appropriate decoherence corrections are included to improve the description of electronic population dynamics.\cite{runeson:2025p154105,granucci:2007p134114}

The electronic energies, nuclear gradients, and nonadiabatic couplings required for the surface-hopping trajectories were evaluated using extended multistate complete active space second-order perturbation theory (XMS-CASPT2).\cite{shiozaki:2011p081106}
XMS-CASPT2 provides a correlated multireference description of excited states and is well suited to the strongly coupled and near-degenerate electronic states encountered along nonadiabatic relaxation pathways.
The use of this high-level electronic structure method therefore enables the trajectories to capture both dynamical electron correlation and changes in electronic character as AcAc evolves through regions of strong nonadiabatic coupling.

To construct the time-resolved XAS spectra, molecular geometries were sampled periodically from the surface-hopping trajectories.
At each geometry, the same active space and state-averaged CASSCF (SA-CASSCF) reference wavefunctions used in the XMS-CASPT2 dynamics were employed to evaluate the MR-ADC carbon K-edge spectrum for the electronic state occupied by the trajectory at that time step.
The resulting spectra were grouped by time delay and, when appropriate, by electronic state before averaging to obtain the simulated transient XAS response.
This procedure establishes a direct connection between the evolving electronic-state populations and nuclear configurations generated by the XMS-CASPT2 surface-hopping dynamics and their corresponding MR-ADC core-excitation spectra.

%%%%%%%%%%%%%%%%%%%%%%%%%%%%%%%%%%%%%%%%%%%%%%%%%%%%%%%%%%%%%%%%%%%%%
\section{Computational Details}
\label{sec:ComputationalDetails}
% 3. Computational Details

The ground-state ($S_0$) equilibrium geometry of acetylacetone (AcAc) was taken from the literature and had been optimized at the CASPT2(10e,8o)/cc-pVDZ level.\cite{severino:2025p30785}
We first benchmarked the main computational parameters entering the MR-ADC simulations against the experimental ground-state carbon K-edge XAS spectrum.
The details of this benchmark are described in \cref{sec:results_and_discussion:gs_xas}. 
In short, our benchmark considered eight active spaces, three second-order MR-ADC approximations [MR-ADC(2), MR-ADC(2)-SX, and MR-ADC(2)-X], and three basis sets (cc-pwCVDZ, aug-cc-pwCVDZ, and cc-pwCVTZ).
For the singlet states, the MR-ADC calculations employed state-averaged CASSCF (SA-CASSCF) reference wavefunctions including the $S_0$, $S_1$, and $S_2$ states.
Based on these tests, MR-ADC(2)/CASSCF(10e,8o)/cc-pwCVDZ was selected for the production TR-XAS simulations.
The CASSCF(10e,8o) active space, which contains the principal $\pi$, $\pi^\ast$, and nonbonding ($n$) orbitals, has also been widely employed in previous multireference studies of AcAc\cite{severino:2025p30785,chen:2006p4434} and provides a balanced description of its low-lying electronic states (Figure S8).
The use of MR-ADC(2) and the cc-pwCVDZ basis set further provides a favorable balance between spectral accuracy and the computational cost required for calculations over large nuclear ensembles.
All MR-ADC calculations employed the spin-free exact-two-component (X2C) Hamiltonian\cite{dyall:2001p9136, liu:2009p031104} to account efficiently for scalar relativistic effects.

To assess the influence of ground-state nuclear motion on the carbon K-edge spectrum, molecular geometry was propagated using the initial conditions generated from quantum thermostat molecular dynamics simulations that were obtained from ref \citenum{chakraborty:2026p1090}.
The trajectories were initialized on $S_0$ and propagated for 50 fs with a 0.5 fs nuclear time step using XMS-CASPT2/SA-CASSCF(10e,8o)/cc-pVDZ.
Energy-based decoherence\cite{granucci:2007p134114} was applied, and nuclear velocities were rescaled following accepted hopping events to conserve the total energy.
A representative subset of 40 geometries was selected from the ensemble at 50 fs, and their MR-ADC(2)/CASSCF(10e,8o)/cc-pwCVDZ spectra were averaged to obtain the vibrationally averaged ground-state XAS spectrum.
All surface-hopping simulations were performed with SHARC-MD\cite{mai:2018pe1370, mai:2023p} interfaced with BAGEL\cite{shiozaki:2018pe1331} for the electronic structure calculations.
Spin--orbit coupling was not included in the singlet-state dynamics.

Single-point excited-state calculations were additionally performed at the $S_1$ and $T_1$ equilibrium geometries optimized at the CASPT2(10e,8o)/cc-pVDZ level obtained from Ref.~\citenum{severino:2025p30785}.
For singlet-state calculations, the CASSCF reference was state averaged over $S_0$, $S_1$, and $S_2$.
For the $T_1$ state, calculations employed the lowest-triplet CASSCF reference.

For the nonadiabatic dynamics following photoexcitation, the initial conditions from ref \citenum{chakraborty:2026p1090} were also utilized. 
These trajectories were initialized on $S_2$ and propagated for 200 fs using the same XMS-CASPT2/SA-CASSCF(10e,8o)/cc-pVDZ electronic structure treatment, 0.5 fs time step, decoherence correction, and energy-conserving velocity rescaling described above.
The use of XMS-CASPT2 provides a correlated multireference description of the closely coupled excited states and nonadiabatic regions sampled during the ultrafast relaxation of AcAc.
Geometries were sampled at 10 fs intervals over the early-time 0--100 fs window and at 20 fs intervals over the later 120--200 fs window.
At each sampling time, 40 trajectories were pseudorandomly selected while preserving the distribution of electronic states and principal structural coordinates of the complete ensemble (Figure S10).
Trajectories that failed to reach 200 fs or had a total energy drift greater than 1.0 eV were omitted from this selection.
Convergence tests showed negligible differences between spectra averaged over 30 and 40 trajectories.
For each sampled geometry, the MR-ADC calculation employed the same active-space definition and corresponding SA-CASSCF reference used in the dynamics, and the carbon K-edge spectrum was evaluated for the electronic state occupied by that trajectory at the selected time step.
The resulting spectra were grouped by time delay and, when required, by electronic state before ensemble averaging to construct the simulated TR-XAS response.

A separate nuclear ensemble was generated to model the long-lived $T_1$ spectrum.
The literature $T_1$ minimum was reoptimized at the CASPT2(10e,8o)/cc-pVDZ level using BAGEL, and the corresponding Hessian was used to generate a 300 K Wigner ensemble following the same procedure employed for $S_0$.
To remove unphysical motion characteristic of Wigner sampling for AcAc, low frequency torsional methyl group and high frequency C--H modes were frozen.\cite{severino:2025p30785}
The resulting $T_1$ trajectories were propagated for 50 fs, after which 40 geometries were randomly selected for MR-ADC calculations and averaging of the triplet-state XAS spectrum.

All single-point and ensemble-averaged spectra were broadened using a Lorentzian representation of the discrete MR-ADC transitions,
\begin{align}
	\label{eq: PES}
	A(\omega) &= -\frac{1}{\pi} \mathrm{Im} \left[ \sum_{\mu}\frac{P_{\mu}}{\omega - \omega_{\mu} + i\eta} \right],
\end{align}
where $\omega_{\mu}$ denotes an MR-ADC core-excitation or valence-ionization energy, $P_{\mu}$ is the corresponding normalized oscillator strength, and $\eta$ controls the phenomenological broadening.
A value of $\eta=0.4$ eV was used for single-point spectra, whereas $\eta=0.2$ eV was employed for vibrationally averaged spectra.
Transient XAS difference spectra were constructed by subtracting the corresponding ground-state $S_0$ spectrum from the excited-state spectrum.
For comparisons involving single-point calculations, the single-point $S_0$ spectrum was used as the reference, whereas ensemble-averaged transient spectra employed the vibrationally averaged $S_0$ reference.
Energy shifts of $-3.5$ and $-3.25$ eV were applied to the single-point and vibrationally averaged MR-ADC(2) spectra, respectively, to align the calculated ground-state bleach maximum with that in the experimental TR-XAS spectra.
A uniform intensity scaling factor was additionally applied when comparing the calculated difference spectra with the experimental signal.

Experimental TR-XAS spectra from Bhattacherjee \textit{et al.}\cite{bhattacherjee:2017p16576}, Haugen \textit{et al.}\cite{haugen:2024p}, and previously reported population dynamics from Severino \textit{et al.}\cite{severino:2025p30785} and Scutelnic \textit{et al.}\cite{scutelnic:2025p5068} were digitized using WebPlotDigitizer.\cite{WebPlotDigitizer}

%%%%%%%%%%%%%%%%%%%%%%%%%%%%%%%%%%%%%%%%%%%%%%%%%%%%%%%%%%%%%%%%%%%%%
% 4. Results
%%%%%%%%%%%%%%%%%%%%%%%%%%%%%%%%%%%%%%%%%%%%%%%%%%%%%%%%%%%%%%%%%%%%%
\section{Results and Discussion}
\label{sec:results_and_discussion} 

\subsection{Ground-State X-ray Absorption Spectrum of AcAc}
\label{sec:results_and_discussion:gs_xas} 

\begin{figure}[t!]
	\centering
	\includegraphics[width=0.95\columnwidth]{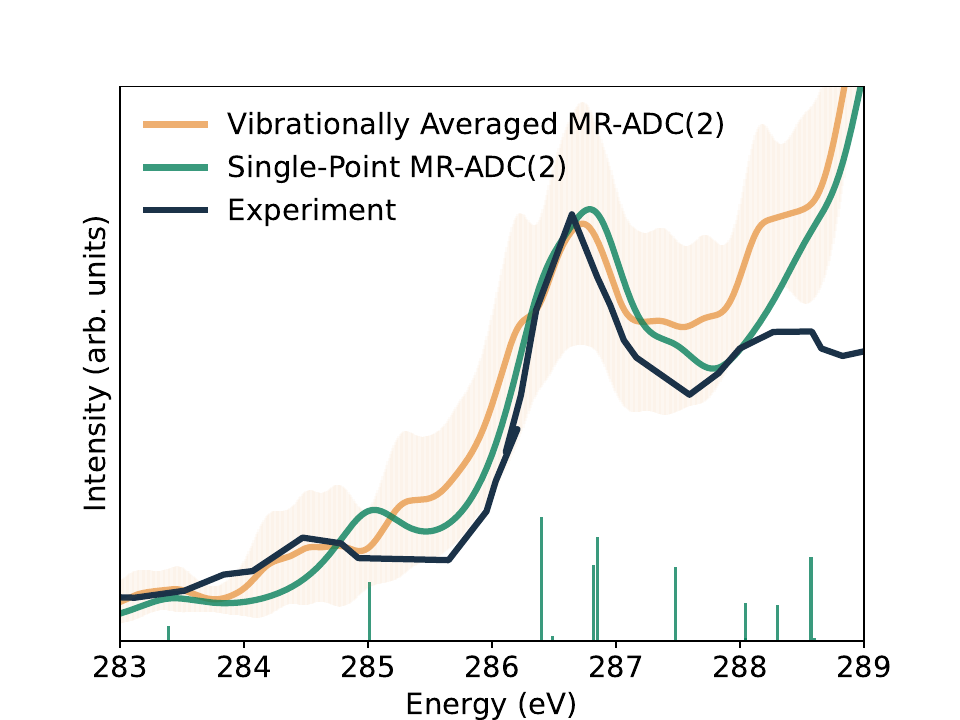}
	\caption{
		Comparison of the experimental ground-state carbon K-edge XAS spectrum of AcAc\cite{bhattacherjee:2017p16576} with single-point and vibrationally averaged MR-ADC(2) spectra.
		The single-point spectrum was computed at the ground-state equilibrium geometry. 
		The vibrationally averaged spectrum was obtained by averaging spectra over geometries sampled from the ground-state nuclear ensemble.
		All calculations employed the MR-ADC(2)/CASSCF(10e,8o)/cc-pwCVDZ protocol.
		The shaded region represents one standard deviation of the intensity distribution at each excitation energy.
	}
	\label{fig:S0_vib_avg}
\end{figure}

Before simulating transient XAS spectra, we first benchmarked the ground-state carbon K-edge XAS spectrum of AcAc.
This benchmark serves three purposes: 
(i) to validate the ability of MR-ADC to reproduce the main experimental ground-state spectral features, 
(ii) to assess the sensitivity of the computed spectrum to active space, basis set, and MR-ADC approximation, and 
(iii) to establish a cost-effective protocol for subsequent excited-state and vibrationally averaged calculations.

Our benchmarking strategy considered three sets of computational parameters.
First, we examined the active-space dependence of the computed spectra using several active spaces with different numbers of electrons and orbitals.
Second, for the selected active space, we compared the results simulated using the cc-pwCVDZ, aug-cc-pwCVDZ, and cc-pwCVTZ basis sets.
Finally, we considered three second-order MR-ADC approximations: MR-ADC(2), MR-ADC(2)-SX, and MR-ADC(2)-X.
All spectra were compared to the experimental ground-state XAS spectrum of AcAc shown in \cref{fig:S0_vib_avg},\cite{bhattacherjee:2017p16576} which exhibits three features at approximately 284.4, 286.6, and 288.2 eV.
The computed excitation energies were systematically shifted to align the position of most intense peak in the MR-ADC spectra with experiment.

At the ground-state equilibrium geometry, the single-point carbon K-edge spectra exhibit a pronounced dependence on the choice of active space (Figures S1--S3).
Among the active spaces considered, 10 electrons in 8 active orbitals (10e,8o) of $\pi$, $\pi^\ast$, and nonbonding ($n$) character provides the most accurate and internally consistent spectra across the different MR-ADC approximations (Figure S8).
This result indicates that these orbitals are essential for describing the low-energy carbon K-edge transitions of AcAc.
The same active space has also been shown in previous multireference studies to provide a reliable description of the AcAc photophysical and photochemical properties.\cite{severino:2025p30785}
We therefore adopted the (10e,8o) active space for all subsequent calculations.

Using the (10e,8o) active space, we next examined the basis-set dependence of the ground-state XAS spectrum at the MR-ADC(2), MR-ADC(2)-SX, and MR-ADC(2)-X levels.
The cc-pwCVDZ, aug-cc-pwCVDZ, and cc-pwCVTZ basis sets yield very similar spectra (Figures S4--S6), and all three reproduce the main experimental features below 289 eV.
Because the larger and more diffuse basis sets do not significantly improve the agreement with experiment in this spectral region, we selected cc-pwCVDZ for the subsequent calculations to reduce computational cost.

With the active space and basis set fixed, MR-ADC(2) and MR-ADC(2)-SX yield nearly identical ground-state XAS spectra.
In contrast, MR-ADC(2)-X produces larger changes in the peak separations and relative intensities, resulting in somewhat poorer agreement with experiment (Figure S7).
On this basis, we selected MR-ADC(2) as the preferred approximation for the subsequent simulations.

\begin{figure}[t!]
	\centering
	\includegraphics[width=1.0\columnwidth]{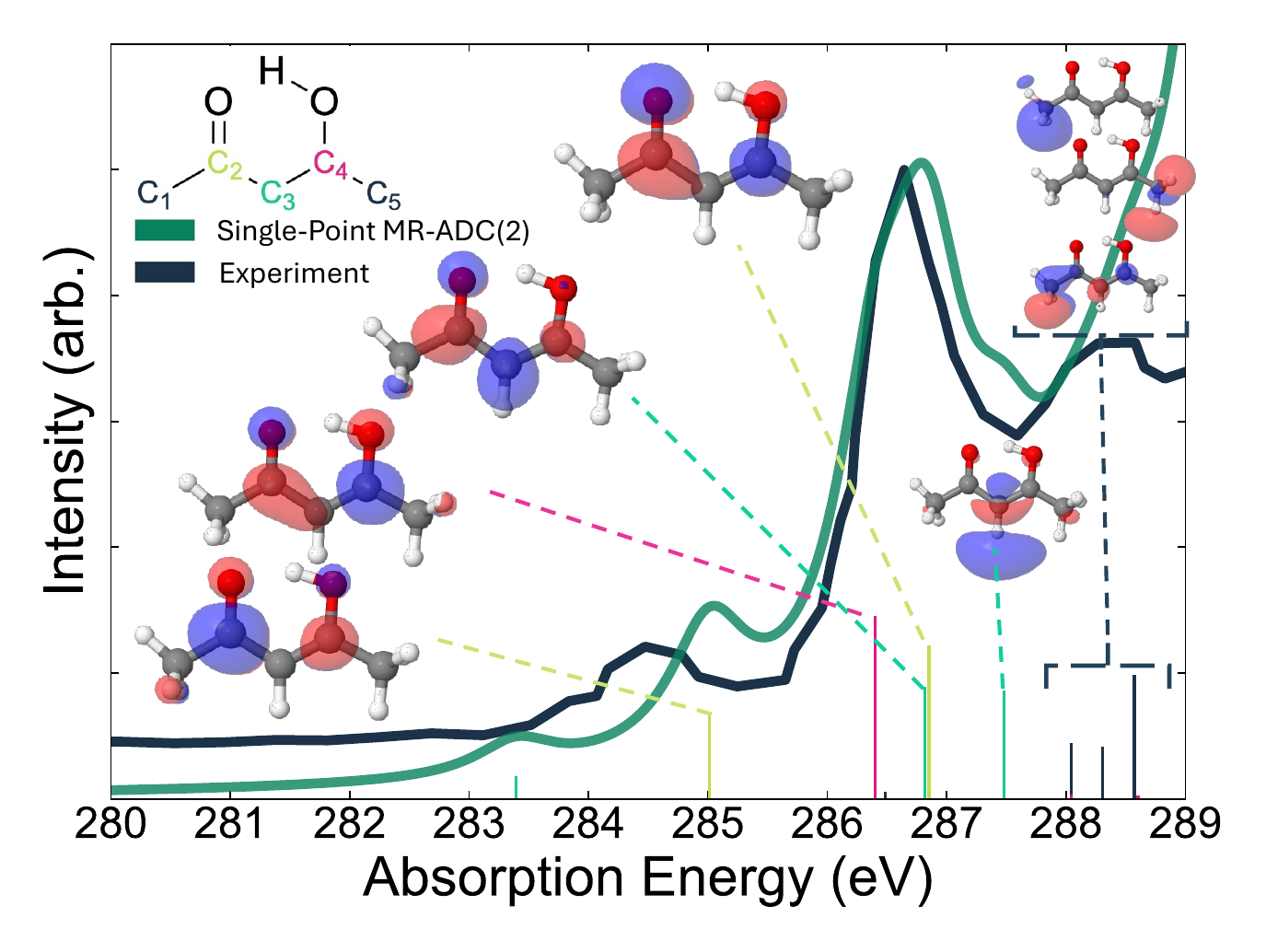}
	\caption{
	Experimental ground-state carbon K-edge XAS spectrum of AcAc\cite{bhattacherjee:2017p16576} compared with the single-point MR-ADC(2)/CASSCF(10e,8o)/cc-pwCVDZ spectrum.
	Vertical sticks indicate the normalized calculated core-excitation energies and oscillator strengths and are color-coded according to the initial carbon 1s orbital (C$_1$--C$_5$).
	}
	\label{fig:s0_xas_nto}
\end{figure}

%In addition to the experimental XAS spectrum, we compared the MR-ADC(2)/CASSCF(10e,8o)/cc-pwCVDZ results with the EOM-CCSD/6-311++G** calculations reported by Faber et al.\cite{faber:2019p144107}. 
%Aside from a uniform energy shift, the MR-ADC(2) spectra show a very good agreement with the EOM-CCSD data for the energy spacing between the first two peaks in the XAS spectrum ($\sim$ 1.6 eV for both methods). 
%While EOM-CCSD underestimates the intensity of the first peak, MR-ADC(2) predicts somewhat higher intensity relative to experiment. 
%As discussed below, these errors are likely affected by missing vibrational effects in these single-point calculations.

In addition to comparison with experiment, we benchmarked the MR-ADC(2)/CASSCF(10e,8o)/cc-pwCVDZ spectrum against the EOM-CCSD/6-311++G** results reported by Faber et al.\cite{faber:2019p144107}
After accounting for an overall energy offset, the two methods show very good agreement in the relative positions of the lowest-energy features, predicting a separation of approximately 1.6 eV between the first two peaks.
While EOM-CCSD underestimates the intensity of the first peak, MR-ADC(2) slightly overestimates it relative to experiment.
As discussed below, these discrepancies are likely influenced in part by the neglect of vibrational nuclear motion in the single-point calculations.

Overall, the single-point MR-ADC(2)/CASSCF(10e,8o)/cc-pwCVDZ spectrum shows a good agreement with reference experimental and theoretical XAS data for the ground electronic state of AcAc (\cref{fig:S0_vib_avg}).
Natural transition orbital (NTO) analysis of the dominant transitions (\cref{fig:s0_xas_nto}) assigns the lowest-energy feature at 284.4 eV primarily to a C~1s $\rightarrow \pi^\ast$ excitation originating from the unprotonated carbonyl carbon, C$_2$.
This assignment differs from the time-dependent density functional theory (TD-DFT) calculations of Bhattacherjee et al.\cite{bhattacherjee:2017p16576} and the EOM-CCSD calculations of Faber et al.,\cite{faber:2019p144107} which attribute the corresponding feature predominantly to excitation from the central carbon atom, C$_3$.
The most intense feature at 286.6 eV comprises C~1s $\rightarrow \pi^\ast$ excitations involving all three central carbon atoms, C$_2$--C$_4$.
The experimental feature at 288.2 eV corresponds to several calculated transitions with substantial C~1s $\rightarrow \sigma^\ast$ and Rydberg character originating primarily from the methyl carbon atoms, C$_1$ and C$_5$.
Apart from the differing assignment of the lowest-energy feature, the MR-ADC(2) interpretation is broadly consistent with the previous TD-DFT and EOM-CCSD studies.
Based on these benchmarks, we selected the MR-ADC(2)/CASSCF(10e,8o)/cc-pwCVDZ protocol for the subsequent transient XAS calculations.

We next assessed the influence of nuclear motion by computing a vibrationally averaged ground-state spectrum from geometries sampled along \textit{ab initio} XMS-CASPT2 dynamics trajectories (see \cref{sec:ComputationalDetails} for details).
For each sampled geometry, the carbon K-edge spectrum was calculated at the MR-ADC(2)/CASSCF(10e,8o)/cc-pwCVDZ level, and the resulting spectra were averaged (\cref{fig:S0_vib_avg}).
Below 287 eV, vibrational averaging produces only modest changes relative to the equilibrium-geometry spectrum, primarily in the form of moderate broadening and small variations in relative peak intensities.
The most noticeable effect occurs for the weak lowest-energy feature, which shifts by approximately 0.5 eV to lower energy and moves into substantially better agreement with experiment.
The overall weak sensitivity of this spectral region to nuclear motion is consistent with the localized character of the dominant low-energy transitions, which primarily involve the conjugated C$_2$--C$_4$ framework and are therefore comparatively insensitive to vibrations of the terminal groups.
In contrast, nuclear averaging has a much stronger effect above 287 eV, where the dominant transitions involve the terminal methyl groups (\cref{fig:s0_xas_nto}).
The resulting redistribution and broadening of intensity significantly improve the agreement with experiment, particularly for the band near 288.2 eV.

These benchmarks demonstrate that MR-ADC reproduces the principal ground-state carbon K-edge features of AcAc using a compact and computationally tractable approach.
The resulting MR-ADC(2)/CASSCF(10e,8o)/cc-pwCVDZ protocol therefore provides a reliable foundation for the transient XAS simulations presented below.

\subsection{Nonadiabatic Dynamics Following $S_2$ Excitation}
\label{sec:results_and_discussion:dynamics} 

 \begin{figure}[t!]
	\centering
	\includegraphics[width=1.0\columnwidth]{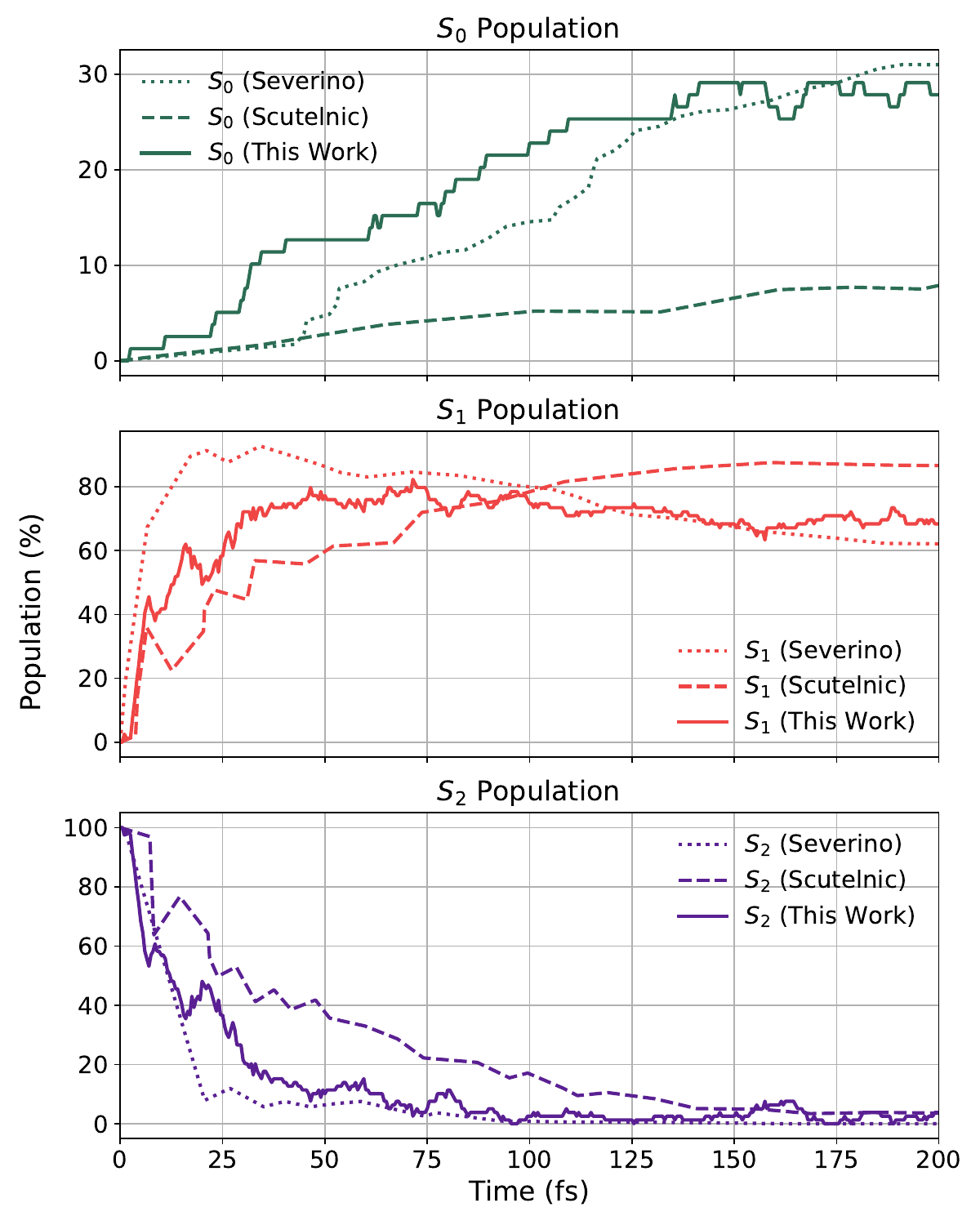}
	\caption{
		Time-dependent populations of the $S_0$, $S_1$, and $S_2$ states extracted from 100 nonadiabatic molecular dynamics trajectories performed at the XMS-CASPT2/CASSCF(10e,8o) level of theory (solid lines).
		For comparison, previously reported population dynamics from Severino \textit{et al.}\cite{severino:2025p30785} and Scutelnic \textit{et al.}\cite{scutelnic:2025p5068} are shown as dotted and dashed lines, respectively.
	}
	\label{fig:pop_dynamics}
\end{figure}

To model the ultrafast electron--nuclear motion underlying the experimental TR-XAS signal, we propagated non-adiabatic molecular dynamics trajectories for 200 fs following excitation to the $S_2$ state.
The simulations were restricted to the singlet manifold ($S_0$, $S_1$, and $S_2$ states), consistent with previous studies showing negligible triplet population during this time window.\cite{severino:2025p30785,scutelnic:2025p5068}

As shown in \cref{fig:pop_dynamics}, the XMS-CASPT2/SA-CASSCF(10e,8o)/cc-pVDZ trajectories exhibit rapid depletion of $S_2$, transient population of $S_1$, and a more gradual rise of $S_0$.
The accompanying nuclear dynamics support a sequential $S_2 \rightarrow S_1 \rightarrow S_0$ relaxation pathway involving excited-state intramolecular hydrogen transfer (ESIHT) and ring opening, in agreement with previous simulations.\cite{severino:2025p30785,scutelnic:2025p5068}
The calculations of Severino et al.,\cite{severino:2025p30785} performed at the same electronic structure level, predict a somewhat faster decay of the $S_2$ population accompanied by a slower rise in the $S_0$ population.
Compared with the SA-CASSCF(10e,9o)/3-21G simulations of Scutelnic et al.,\cite{scutelnic:2025p5068} our trajectories predict a more rapid depletion of the $S_2$ population but a broadly similar evolution of the $S_1$ population.
Despite these quantitative differences, all three studies support the same mechanistic picture of rapid, nuclear-motion-driven relaxation following $S_2$ excitation.

These nonadiabatic trajectories provide the time-dependent nuclear ensembles required to connect the evolving excited-state populations and molecular geometries with the measured TR-XAS response.
In the following sections, we analyze the transient carbon K-edge XAS spectra to determine how coupled electronic and nuclear dynamics govern the observed spectral evolution.

\subsection{Early-Time (20--100 fs) TR-XAS Spectra}
\label{sec:results_and_discussion:20_100fs} 

\begin{figure}[t!]
	\centering
	\includegraphics[width=1.0\columnwidth]{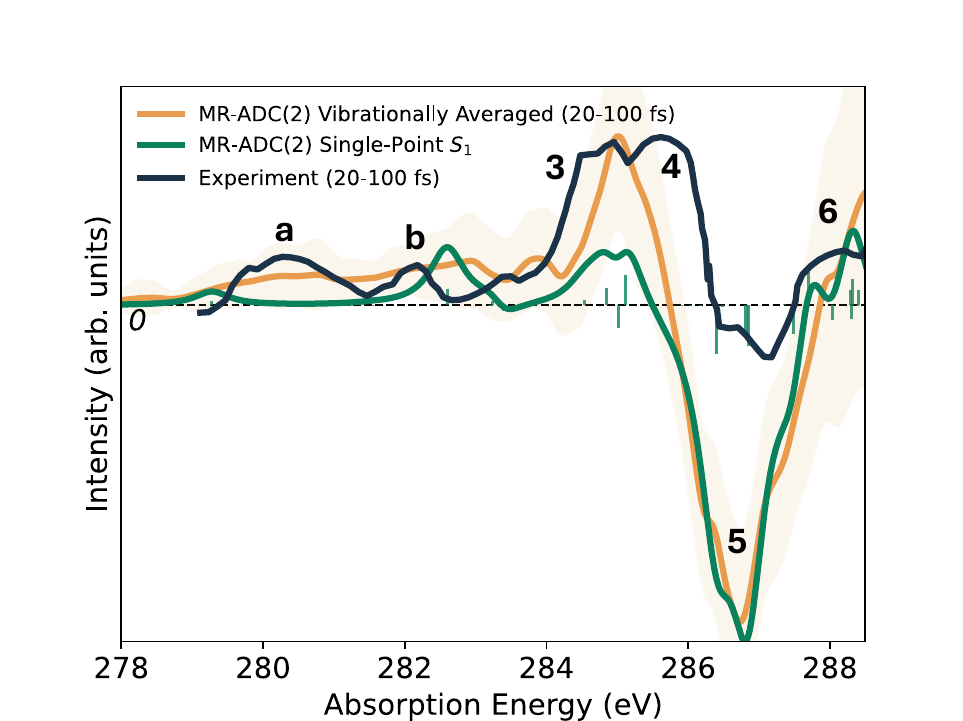}
	\caption{
		Comparison of the experimental\cite{bhattacherjee:2017p16576} and calculated transient XAS spectra averaged over the 20--100 fs time window.
		All simulated spectra were computed using the MR-ADC(2)/CASSCF(10e,8o)/cc-pwCVDZ protocol, see \cref{sec:ComputationalDetails} for additional details.
		The shaded region represents one standard deviation of the calculated intensity at each excitation energy in the vibrationally averaged spectrum.
	}
	\label{fig:20_100fs}
\end{figure}

\begin{figure}[t!]
	\centering
	\includegraphics[width=1.0\columnwidth]{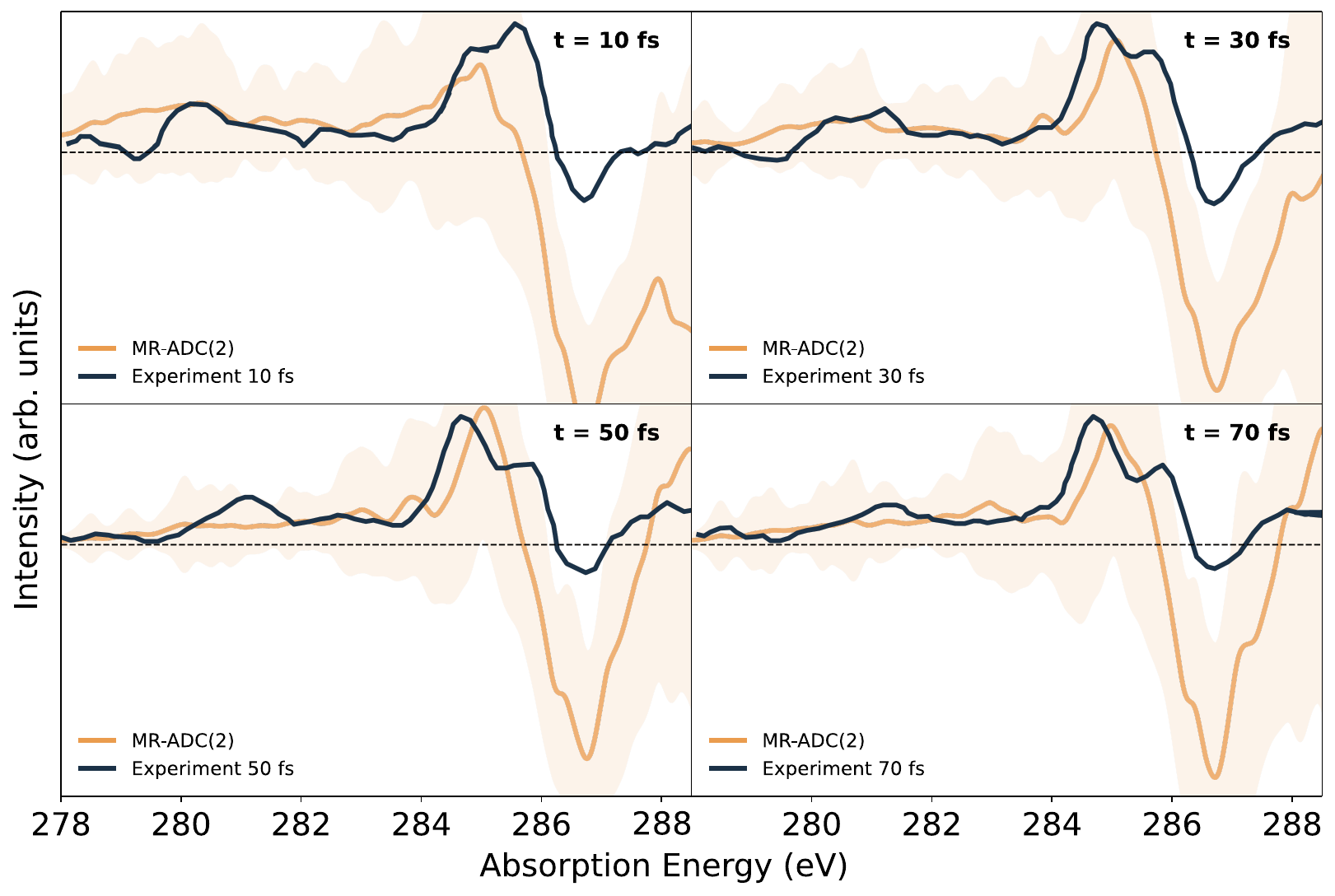}
	\caption{
		Comparison of the experimental\cite{haugen:2024p} and calculated transient XAS spectra of AcAc for the 10 -- 70 fs time window following photoexcitation.
		All simulated spectra were computed using the MR-ADC(2)/CASSCF(10e,8o)/cc-pwCVDZ protocol, see \cref{sec:ComputationalDetails} for additional details.
		The shaded region represents one standard deviation of the calculated intensity at each excitation energy in the vibrationally averaged spectrum.
		}
	\label{fig:70fs_haugen}
\end{figure}

We first evaluate the calculated early-time TR-XAS response against two complementary experimental measurements.
Bhattacherjee et al.\cite{bhattacherjee:2017p16576} reported a spectrum averaged over the 20--100 fs interval with an instrument response time of 90 fs (\cref{fig:20_100fs}), whereas Haugen et al.\cite{haugen:2024p} measured the spectral evolution within the first 70 fs with a 20 fs instrument response time (\cref{fig:70fs_haugen}).

The 20--100 fs spectrum of Bhattacherjee et al.\cite{bhattacherjee:2017p16576} exhibits weak positive absorption between 279.5 and 282.5 eV (peaks a and b), followed by a much stronger excited-state absorption envelope spanning 284--286 eV, with partially resolved maxima near 284.5 and 285.6 eV (peaks 3 and 4).
This positive signal is followed by a pronounced ground-state bleach centered near 286.8 eV (peak 5).
At higher energies, the transient signal becomes positive again above approximately 287.5 eV and reaches a maximum near 288.8 eV (peak 6).
To model the same time window, we sampled 40 trajectories pseudorandomly at 10 fs intervals between 20 and 100 fs and averaged the corresponding MR-ADC(2) spectra.
As shown in Figure S10, the sampled subset closely reproduces the distributions of the principal structural coordinates and electronic-state populations of the complete trajectory ensemble.
Moreover, increasing the sample size from 30 to 40 trajectories produces negligible changes in the averaged spectrum, indicating that the calculated spectral profile is well converged with respect to trajectory sampling.

The vibrationally averaged MR-ADC(2) spectrum qualitatively reproduces the principal experimental features (\cref{fig:20_100fs}), including the low-energy excited-state absorption (279.5--282.5 eV), the broader positive envelope between 284 and 286 eV, the ground-state bleach near 286.8 eV, and the excited-state absorption at 288.8 eV.
Although the magnitude of the bleach is substantially overestimated, the ensemble-averaged spectrum (orange curve) provides a markedly better description of the experimental features than the single-point spectrum evaluated at the stationary $S_1$ geometry (green curve).
The improvement is particularly evident in the low-energy absorption region (peaks a and b) and in the 284--286 eV envelope (peaks 3 and 4), which broadens and becomes more intense relative to other features.
These differences demonstrate that non-equilibrium nuclear motion plays a central role in shaping the early-time transient spectrum.

\begin{figure}[t!]
	\centering
	\includegraphics[width=1.0\columnwidth]{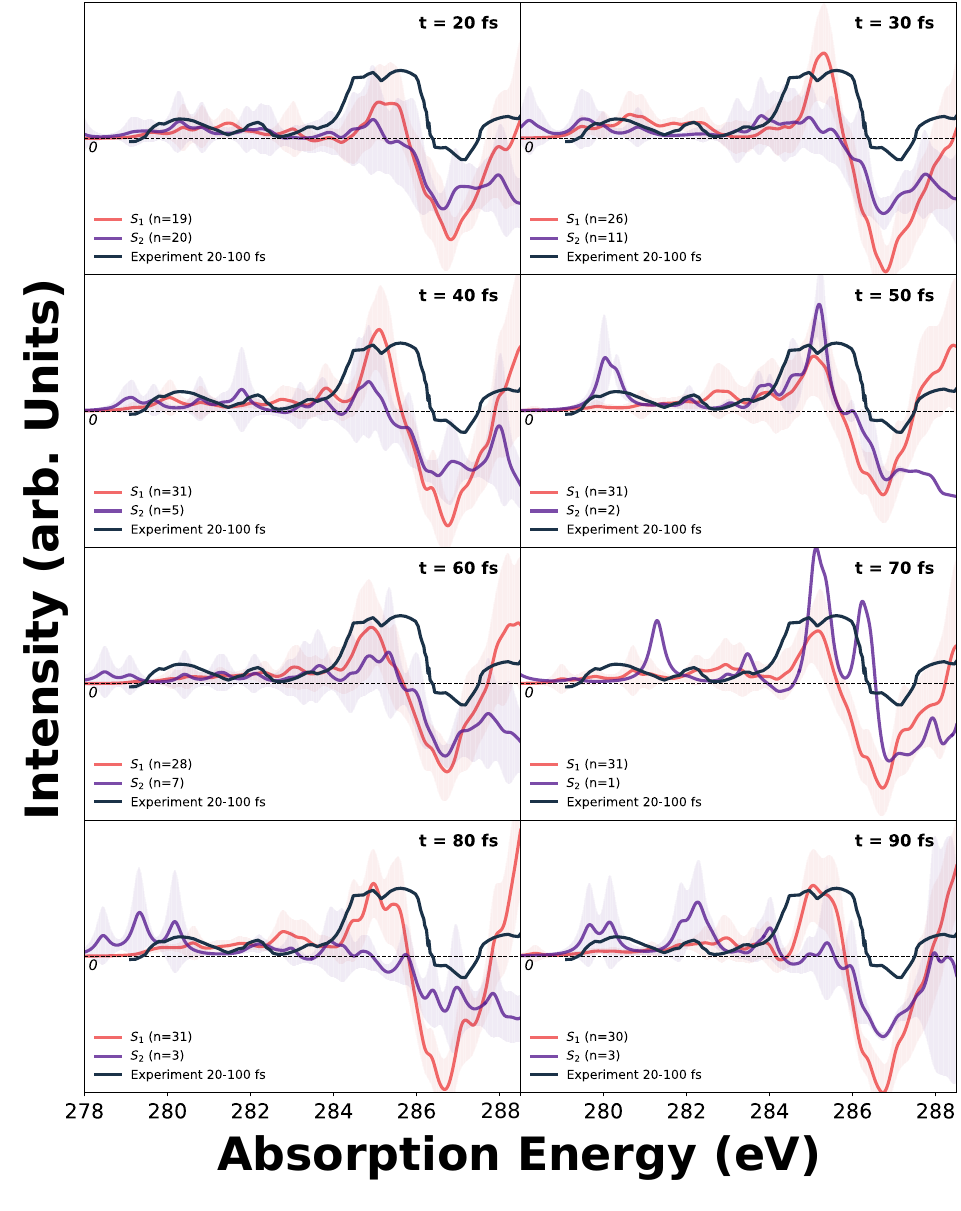}
	\caption{
		State-resolved transient XAS spectra obtained by averaging over 40 sampled non-adiabatic trajectories at selected time delays $t$ following excitation to the $S_2$ state.
		All spectra were computed at the MR-ADC(2)/CASSCF(10e,8o)/cc-pwCVDZ level.
		The numbers in the legends indicate the number of trajectories contributing to each state-specific average.
		The corresponding experimental spectrum averaged over 20--100 fs is overlaid in each panel.
		Shaded regions represent one standard deviation of the calculated intensity within each state-specific ensemble.
		Contributions from trajectories occupying the $S_0$ state are omitted for clarity because their XAS signals largely overlap with the ground-state bleach.
	}
	\label{fig:time_series_early}
\end{figure}

To determine how individual time delays contribute to the 20--100 fs average, we next examined state-resolved spectra computed at 10 fs intervals (\cref{fig:time_series_early}).
At 20 fs, the sampled ensemble contains nearly equal populations of both $S_2$ and $S_1$, whereas $S_1$ becomes the dominant excited state at later delays.
Despite this population transfer, the $S_1$ and $S_2$ ensembles exhibit broadly similar transient line shapes.
Both states contribute weak absorption in the 279.5--282.5 eV region, which is most prominent during the first 60 fs and diminishes substantially at later times.
Stronger absorption from $S_1$ and $S_2$ between 284 and 286 eV persists throughout the 20--100 fs interval, with the relative intensities of spectral features varying among the sampled time delays.
The qualitative similarity of the state-resolved spectra indicates that the spectral evolution cannot be interpreted solely in terms of population transfer from $S_2$ to $S_1$.
Instead, much of the variation arises from the broad and continuously evolving distribution of nuclear geometries sampled within each electronic state.

\begin{figure}[t!]
	\centering
	\includegraphics[width=1.0\columnwidth]{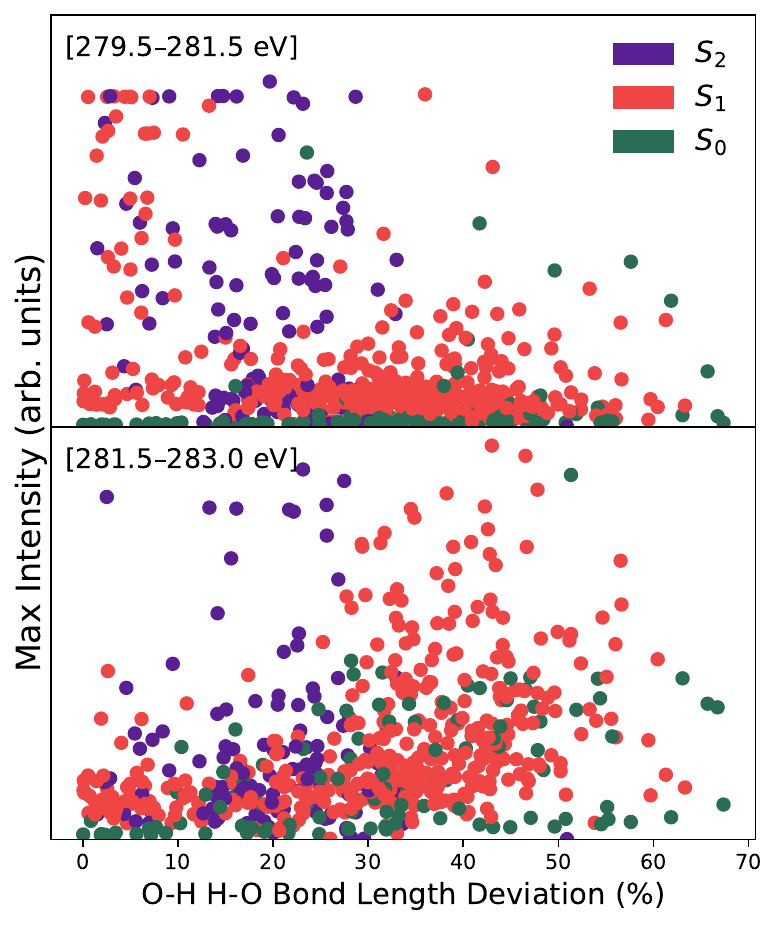}
	\caption{
		Maximum calculated transient XAS intensity in the 279.5--281.5 eV and 281.5--283 eV region as a function of the relative O--H/H--O bond-length deviation for all 560 computed MR-ADC spectra (20 -- 200 fs). 
		Small deviations correspond to approximately $C_{2v}$-symmetric structures in which the enolic proton is nearly equidistant from the two oxygen atoms.
		Larger deviations correspond to asymmetric, $C_s$-like enol structures.
	}
	\label{fig:proton_symmetry}
\end{figure}

To determine the structural origin of the low-energy 279.5--282.5 eV signal (peaks a and b in \cref{fig:20_100fs}), we examined the individual spectra and molecular geometries contributing appreciable intensity in this region.
The strongest absorption between 279.5 and 281.5 eV (peak a) is predominantly associated with structures in which the enolic proton is nearly equidistant from the two oxygen atoms.
These structures correspond to approximately $C_{2v}$-symmetric, ring-like proton-sharing configurations transiently accessed during excited-state intramolecular hydrogen transfer.
As shown in \cref{fig:proton_symmetry}, the largest intensities in this energy range occur at small O--H/H--O bond-length deviations, whereas strongly asymmetric, $C_s$-like geometries generally contribute little absorption.
Both the $S_1$ and $S_2$ ensembles contribute to this feature, indicating that it is governed primarily by the proton-sharing geometry rather than by the identity of the electronic state.
The 279.5--281.5 eV signal (peak a) therefore provides a sensitive spectroscopic signature of the short-lived proton-sharing configurations sampled during the earliest stages of AcAc photorelaxation.
In contrast, the absorption between 281.5 and 283 eV (peak b) exhibits a much weaker dependence on the proton-sharing coordinate (\cref{fig:proton_symmetry}), with symmetric and asymmetric geometries contributing comparable intensities.

\begin{figure}[t!]
	\centering
	\includegraphics[width=1.0\columnwidth]{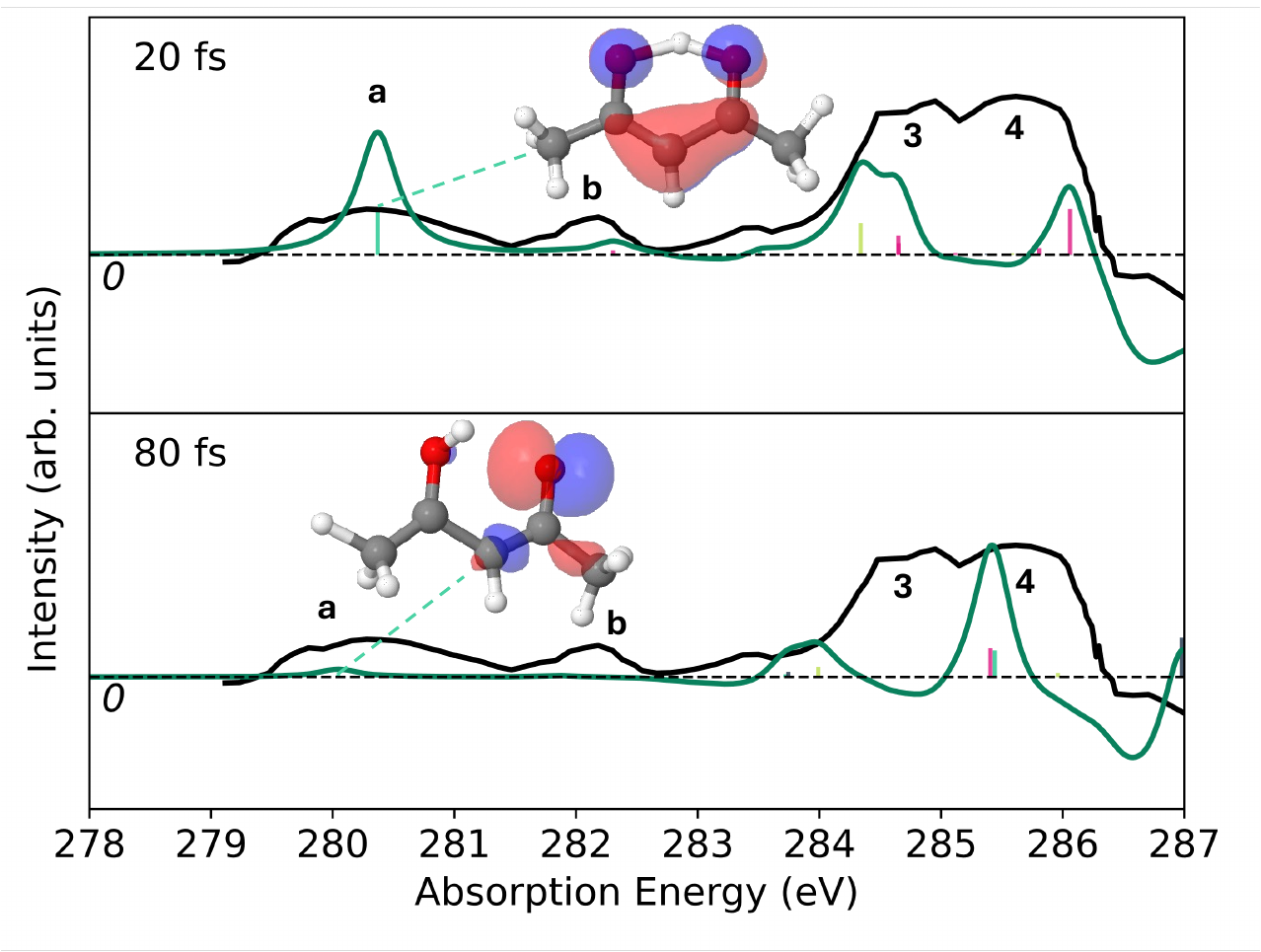}
	\caption{
		Representative non-equillibrium structures and TR-XAS spectra calculated from a single trajectory at 20 and 80 fs. 
		Both TR-XAS spectra are calculated from the $S_1$ state with natural transition orbitals included for comparison between two geometries. 
		The carbon core-orbitals are color-labeled $C_1$-$C_5$ following the convention in \cref{fig:s0_xas_nto}.}
	\label{fig:proton_sharing_nto}
\end{figure}

The NTO analysis provides a microscopic explanation for the strong dependence of peak a on the proton-sharing coordinate.
For the approximately $C_{2v}$-symmetric geometry sampled at 20 fs, the NTO contributing to peak a is strongly localized on C$_3$, resulting in substantial spatial overlap with the C$_3$ 1s core orbital and enhancing the intensity of the C$_3$ 1s $\rightarrow \pi^\ast$ transition (\cref{fig:proton_sharing_nto}).
In contrast, proton localization on one oxygen in the asymmetric geometry sampled at 80 fs redistributes the NTO toward the adjacent oxygen center.
The reduced overlap of this NTO with the C$_3$ 1s orbital substantially decreases the corresponding transition intensity and suppresses peak a.
This analysis shows that the intensity of the low-energy feature is controlled not only by the transition energy but also by geometry-dependent changes in the localization of the accepting $\pi^\ast$ orbital.

The intense absorption between 284 and 286 eV (peaks 3 and 4 in \cref{fig:20_100fs}) is strongly modulated by the evolving nuclear geometry.
The underlying excited-state signal comprises C~1s excitations whose energies and oscillator strengths vary with bond alternation along the central C$_2$--C$_4$ framework and with migration of the enolic proton between the two oxygen atoms.
These coupled motions reorganize the C--C and C--O bonding pattern, redistribute the $\pi$-electron density, and continuously modify the local electronic environments of the central carbon atoms.
As a result, the corresponding core-excitation energies and intensities evolve across the 283.5--286 eV region as the molecule samples different stages of proton transfer and skeletal reorganization (\cref{fig:time_series_early}).
Whereas the experimental spectrum resolves this region into a distinct doublet-like feature, the vibrationally averaged MR-ADC(2) spectrum is dominated by a single broad maximum located near experimental peak 3.
The missing intensity near peak 4 likely reflects, at least in part, an overestimation of the ground-state bleach in the MR-ADC(2) difference spectrum, which suppresses positive excited-state absorption in this energy range.
A similar behavior was reported by Tsuru et al.,\cite{tsuru:2021p024101} whose low-spin open-shell reference EOM-CCSD calculations for the $S_2$ state at the $S_1$ equilibrium geometry likewise failed to recover peak 4 and overestimated the magnitude of the ground-state bleach.

The higher temporal resolution of the TR-XAS measurements reported by Haugen et al.\cite{haugen:2024p} provides a more stringent test of the calculated spectral evolution during the first 70 fs (\cref{fig:70fs_haugen}).
To account for the 20 fs instrument response time, the calculated spectrum at each experimental delay ($t=10$, 30, 50, and 70 fs) was obtained by averaging spectra evaluated at $t$ and $t\pm10$ fs, yielding an ensemble of 120 spectra for each delay.
The calculations reproduce the pronounced redistribution of the weak low-energy absorption from approximately 279.5--281.5 eV at 10 fs toward higher energies as the dynamics progress.
Agreement is particularly good at 10 and 30 fs, where both theory and experiment show a rapid loss of the lowest-energy intensity accompanied by the buildup of weaker absorption in the 282--284 eV region.
This evolution is consistent with the structural picture described above: as the short-lived, nearly symmetric proton-sharing configurations disappear, their characteristic low-energy absorption is suppressed and spectral weight shifts toward transitions associated with more asymmetric geometries.

The calculations also reproduce the progressive growth of peak 3 near 285 eV over the 10--70 fs interval.
A concomitant reduction of intensity in the region associated with peak 4 is also present in the calculated spectra.
This trend is less apparent because the positive excited-state absorption in this energy range increasingly overlaps with the ground-state bleach, which partially obscures the decay of the underlying feature.

Taken together, the time-averaged measurements of Bhattacherjee et al.\cite{bhattacherjee:2017p16576} and the higher-time-resolution spectra of Haugen et al.\cite{haugen:2024p} provide a consistent picture of the first 100 fs of AcAc photorelaxation.
The calculations reproduce both the overall 20--100 fs spectral envelope and key time-dependent trends resolved within the first 70 fs, including the rapid disappearance and blue shift of the low-energy absorption and the redistribution of intensity across the 284--286 eV region.
These observations support a picture in which the transient XAS response arises from a heterogeneous, continuously evolving $S_2$/$S_1$ ensemble undergoing proton transfer, bond alternation, and ring-opening motion.
In particular, the low-energy feature provides a sensitive marker of short-lived proton-sharing structures, while the evolution of peaks 3 and 4 reflects the accompanying reorganization of the central carbon framework and its overlap with the ground-state bleach.

\subsection{Later-Time (120--200 fs) TR-XAS Spectra}
\label{sec:results_and_discussion:120_200fs} 

\begin{figure}[t!]
	\centering
	\includegraphics[width=1.0\columnwidth]{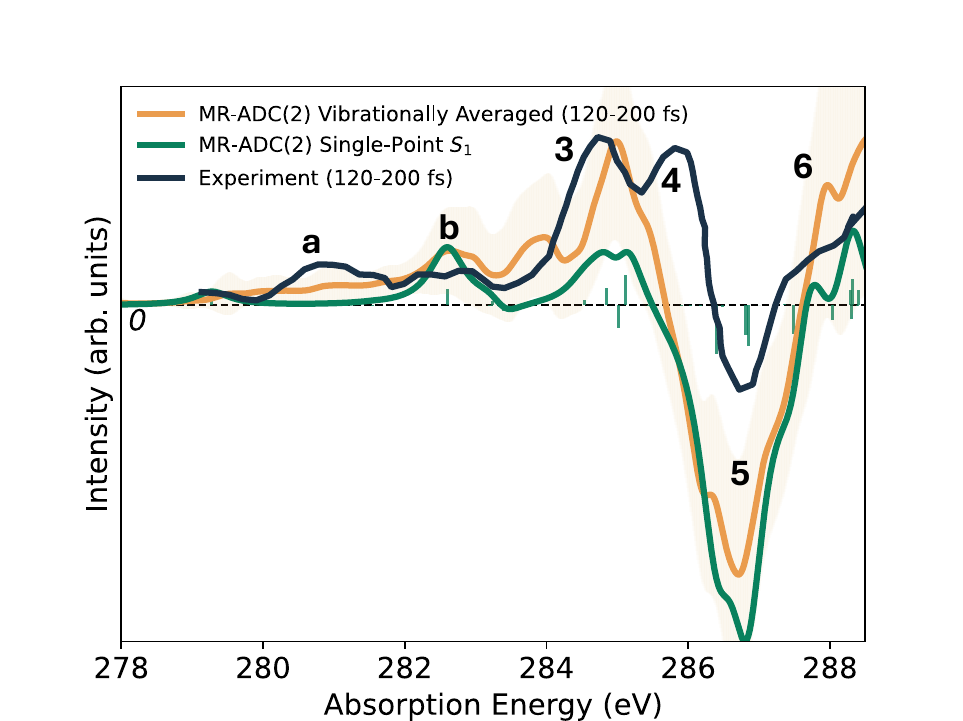}
	\caption{
		Comparison of the experimental\cite{bhattacherjee:2017p16576} and calculated transient XAS spectra averaged over the 120--200 fs time window.
		All simulated spectra were computed using the MR-ADC(2)/CASSCF(10e,8o)/cc-pwCVDZ protocol, see \cref{sec:ComputationalDetails} for additional details.
		The shaded region represents one standard deviation of the calculated intensity at each excitation energy in the vibrationally averaged spectrum. 
	}
	\label{fig:120_200fs}
\end{figure}

\begin{figure}[t!]
	\centering
	\includegraphics[width=1.0\columnwidth]{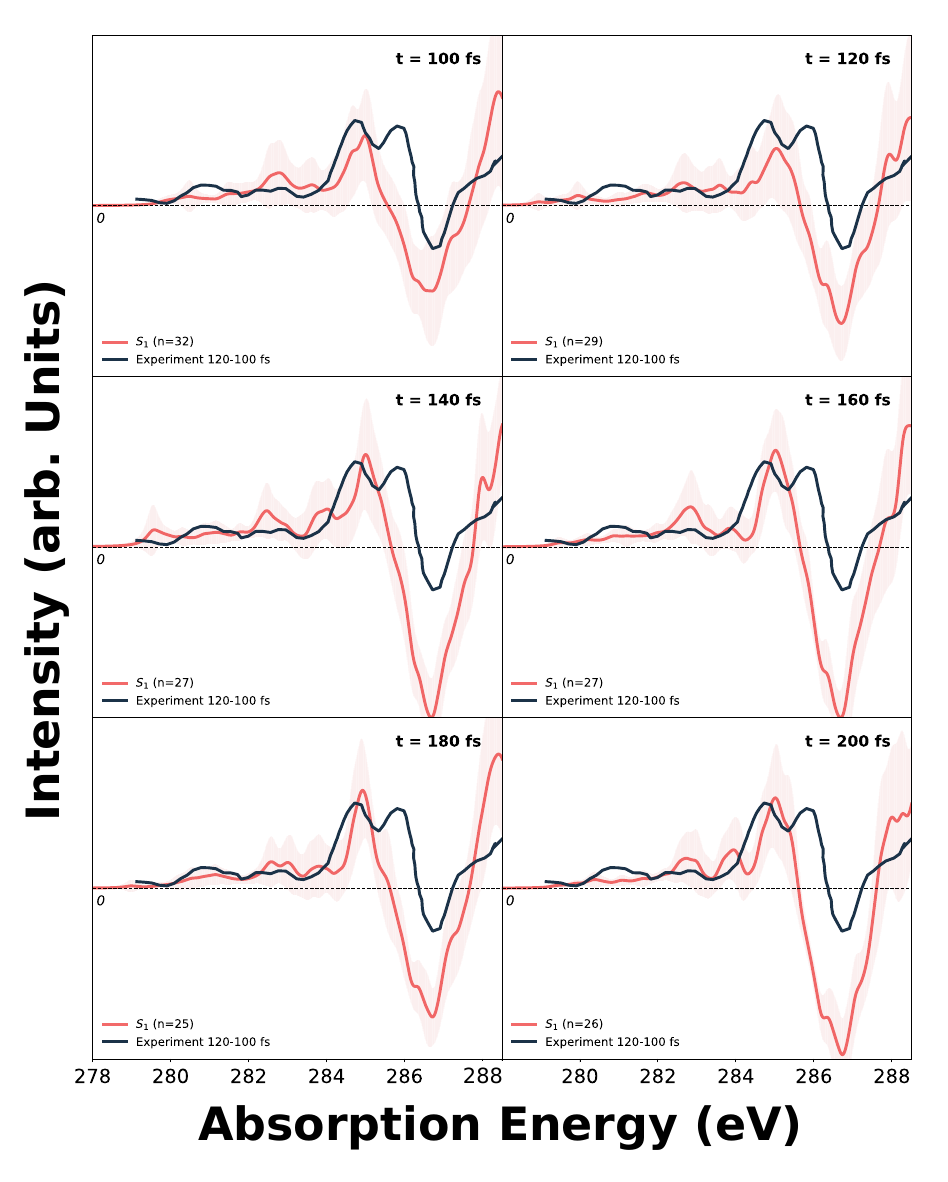}
	\caption{
		State-resolved transient XAS spectra obtained by averaging over 40 sampled non-adiabatic trajectories at selected time delays $t$ following excitation to the $S_2$ state.
		All spectra were computed at the MR-ADC(2)/CASSCF(10e,8o)/cc-pwCVDZ level.
		The numbers in the legends indicate the number of trajectories contributing to each state-specific average.
		The corresponding experimental spectrum is overlaid in each panel.
		Shaded regions represent one standard deviation of the calculated intensity within each state-specific ensemble.
		Contributions from trajectories occupying the $S_0$ state are omitted for clarity because their XAS signals largely overlap with the ground-state bleach.
	}
	\label{fig:time_series_late}
\end{figure}

We next turn to the 120--200 fs transient XAS spectrum (\cref{fig:120_200fs}) reported by Bhattacherjee et al.,\cite{bhattacherjee:2017p16576} which exhibits several subtle but systematic changes relative to the earlier 20--100 fs spectrum (\cref{fig:20_100fs}).
As observed in the experimental data by Haugen et al.,\cite{haugen:2024p} from 20--100 fs to 120--200 fs the weak low-energy absorption persists but shifts upward by approximately 1 eV, from 279.5--282.5 eV to 280.5--283.5 eV, and develops a more structured profile with several weak maxima (peaks a and b in \cref{fig:120_200fs}).
The stronger excited-state absorption between 284 and 286 eV also gains intensity and resolves into a more pronounced doublet with maxima near 284.7 and 285.9 eV.
At the same time, the ground-state bleach near 286.6 eV becomes more intense.
Above the bleach, the positive excited-state signal also increases, while its maximum shifts from approximately 288.8 eV at early times to 288.4 eV in the 120--200 fs spectrum.

The vibrationally averaged MR-ADC(2) spectrum qualitatively captures these changes (\cref{fig:120_200fs}).
In particular, the weak low-energy features shift to higher excitation energies, in agreement with experiment, and the 284--286 eV absorption becomes stronger and develops a more clearly resolved two-peak structure.
The calculated doublet is more asymmetric than the experimental feature and is displaced to lower energy by approximately 0.5 eV.
In contrast to experiment, the calculated ground-state bleach becomes somewhat weaker at later times because a growing fraction of the trajectories has returned to $S_0$ through nonradiative relaxation (\cref{fig:pop_dynamics}), reducing the net depletion of the ground-state contribution.
The positive absorption above the bleach also increases in intensity and shifts by approximately 0.1 eV toward the experimental maximum.
Despite the remaining discrepancies in the doublet profile and bleach amplitude, the calculations reproduce the principal trends in the spectral evolution from 20--100 to 120--200 fs.

The state-resolved spectra provide further insight into the origin of these changes (\cref{fig:time_series_late}).
Between 120 and 200 fs, approximately 70\% of the trajectories occupy $S_1$, less than 5\% remain in $S_2$, and the remainder have returned to $S_0$ (\cref{fig:pop_dynamics}).
The excited-state absorption in this time window is therefore dominated by the evolving $S_1$ ensemble, while the growing $S_0$ population and the ground-state bleach also shape the overall transient difference spectrum.
As in the 20--100 fs interval, spectra sampled at individual time delays exhibit substantial broadening and fluctuations in relative peak intensities.
These variations reflect the heterogeneous nuclear configurations generated by bond alternation, proton transfer, and progressive ring opening, rather than changes in electronic-state populations alone.
The attenuation and blue shift of the low-energy signal (peaks a and b) are likewise consistent with a reduced contribution from the nearly symmetric proton-sharing structures that dominate the earliest-time response.

Our interpretation of the 284--286 eV doublet (peaks 3 and 4) differs substantially from the assignment proposed by Bhattacherjee et al.\cite{bhattacherjee:2017p16576}
Based on TD-DFT spectra evaluated at a limited set of stationary geometries, that study associated peak 4 near 286 eV primarily with $S_2$ population remaining in the Franck--Condon region, whereas peak 3 near 285 eV was attributed to molecules approaching the relaxed $S_1$ or $S_2$ minima.
The trajectory-resolved spectra in \cref{fig:time_series_early,fig:time_series_late} support a different picture.
In particular, during the 120--200 fs interval the $S_2$ population is already small and contributes only weakly to the transient spectrum, making it unlikely that peak 4 originates predominantly from Franck--Condon-like $S_2$ structures.
Instead, our calculations indicate that the 284--286 eV region is formed by a broad manifold of C~1s excitations into low-lying singly occupied $\pi$ and unoccupied $\pi^\ast$ orbitals whose energies and oscillator strengths evolve continuously with proton transfer, bond alternation, and ring-opening motion.
The two apparent maxima therefore emerge from the combined effects of structural heterogeneity, time-dependent redistribution of excited-state intensity, and overlap with the ground-state bleach, rather than from spectroscopically distinct Franck--Condon and relaxed-state species.
A similar picture emerged from the nonorthogonal configuration interaction study of Arias-Martinez et al.,\cite{arias-martinez:2024p752} which found substantial absorption in the 284--286 eV region from both the $S_2$ ($\pi\pi^\ast$) and $S_1$ ($n\pi^\ast$) states, with the dominant spectral structure evolving as population transfers between them.

Overall, the changes between the 20--100 and 120--200 fs spectra reflect the transition from a mixed $S_2$/$S_1$ ensemble dominated by early proton-transfer dynamics to a predominantly $S_1$ ensemble undergoing continued skeletal relaxation and partial return to $S_0$.
The calculated spectral evolution supports an interpretation in which the later-time doublet and accompanying broad features are governed by the combined effects of electronic-state character and a heterogeneous distribution of nuclear geometries.

\subsection{Long-Time (7--10 ps) Triplet-State TR-XAS Spectra}
\label{sec:results_and_discussion:triplet} 

\begin{figure}[t!]
	\centering
	\includegraphics[width=1.0\columnwidth]{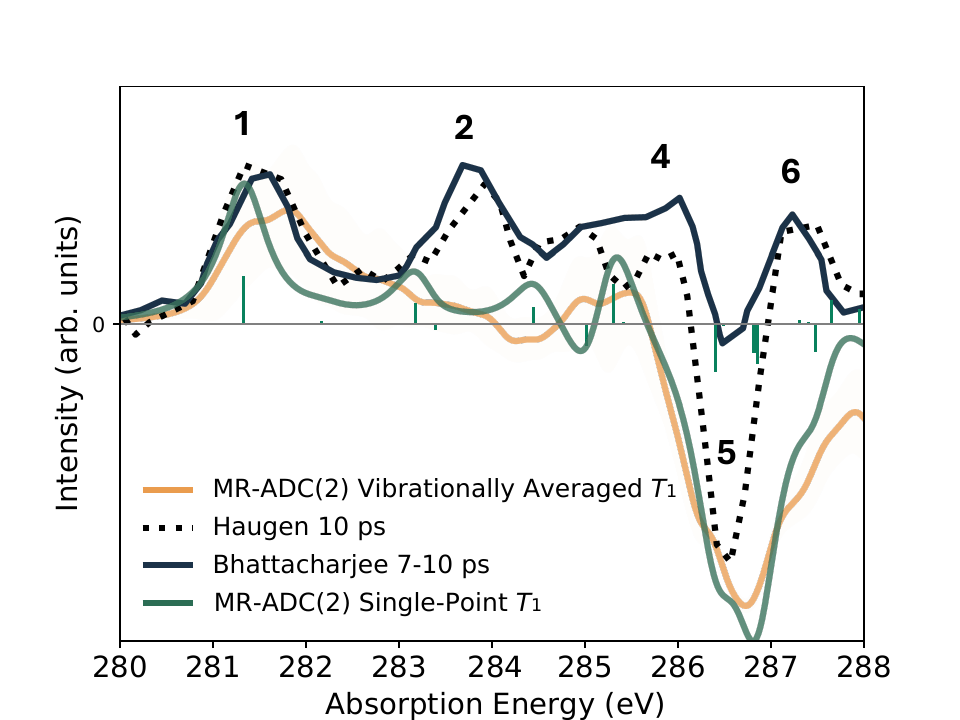}
	\caption{
	Comparison of the experimental TR-XAS difference spectrum of AcAc averaged over 7--10 ps\cite{bhattacherjee:2017p16576} with the calculated triplet-state spectra.
	All calculated spectra were obtained using the MR-ADC(2)/CASSCF(10e,8o)/cc-pwCVDZ protocol.
	Additional computational details are provided in \cref{sec:ComputationalDetails}.
	The shaded region surrounding the vibrationally averaged spectrum indicates one standard deviation of the calculated intensity at each excitation energy.
	}
	\label{fig:T1_vib_avg}
\end{figure}

\begin{figure}[t!]
	\centering
	\includegraphics[width=1.0\columnwidth]{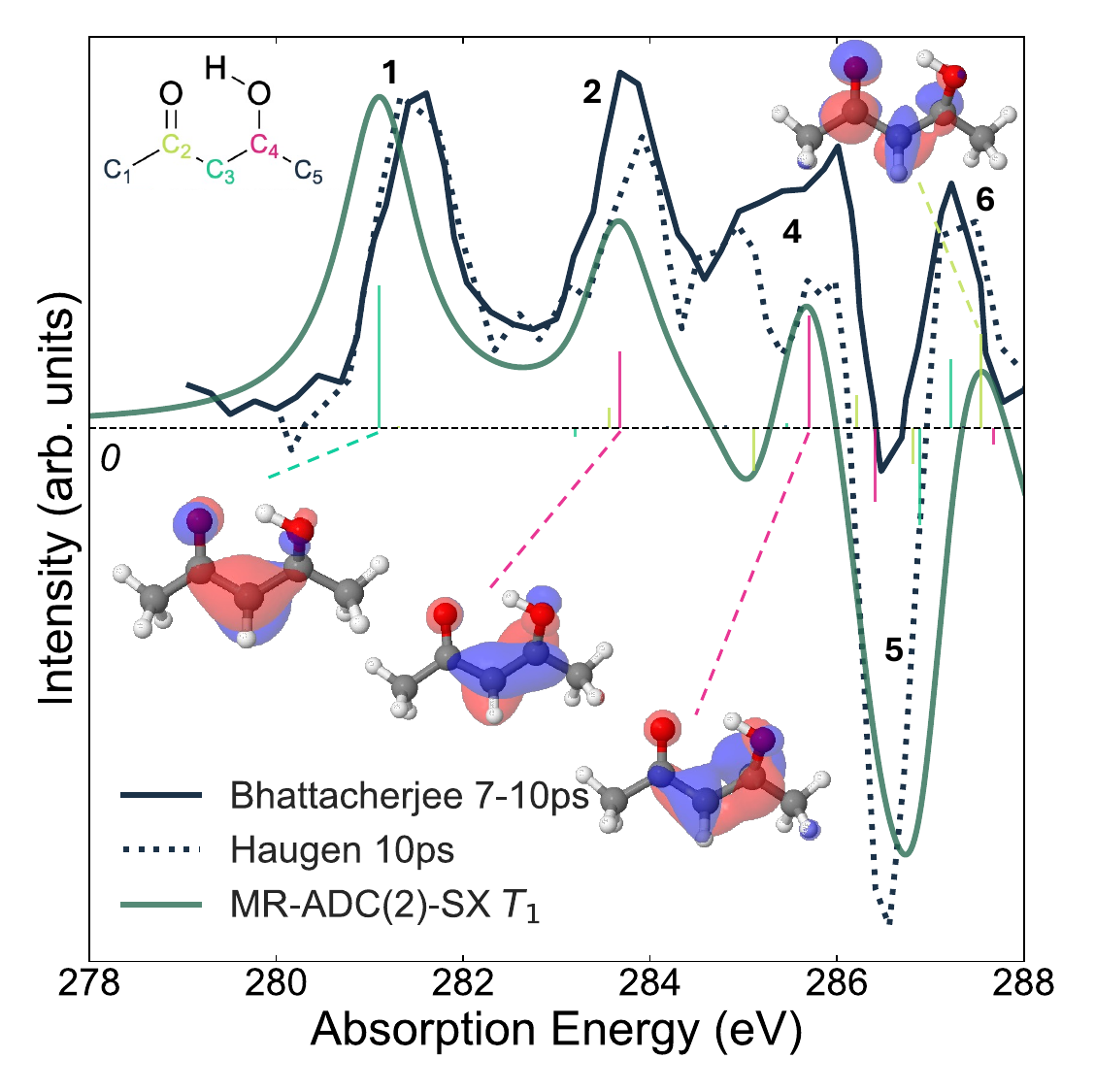}
	\caption{
	Experimental transient XAS difference spectrum of AcAc averaged over 7--10 ps\cite{bhattacherjee:2017p16576} compared with the single-point difference spectrum computed using the MR-ADC(2)-SX/CASSCF(10e,8o)/cc-pwCVDZ protocol.	
	The calculated transitions are color-coded according to the initial carbon 1s orbital, with the five carbon sites labeled C$_1$--C$_5$.
	The calculated spectrum was shifted by $-2.0$ eV to align it with experiment.
		}
	\label{fig:nto_triplet}
\end{figure}

Following relaxation to $S_1$, AcAc undergoes vibrationally activated intersystem crossing to the long-lived $T_1$ state with an experimental time constant of approximately 1.5 ps.\cite{bhattacherjee:2017p16576,severino:2025p30785}
The long-time TR-XAS spectra reported by Bhattacherjee et al.\cite{bhattacherjee:2017p16576} over 7--10 ps and by Haugen et al.\cite{haugen:2024p} at 10 ps therefore predominantly reflect the $T_1$ population (\cref{fig:T1_vib_avg}).
Both measurements exhibit the same principal low-energy signatures, with strong excited-state absorption near 281.4 and 283.8 eV (peaks 1 and 2), followed by a ground-state bleach near 286.6 eV (peak 5) and recovery of positive absorption above 287 eV (peak 6).
The largest differences occur in the 285--287 eV region.
In the Bhattacherjee spectrum, peak 4 appears as a broad feature centered near 285.5 eV, whereas the higher-resolution Haugen spectrum reveals additional structure in this region and a substantially deeper ground-state bleach.
Despite these differences in the higher-energy line shape, the close correspondence of peaks 1, 2, and 6 provides a consistent experimental characterization of the long-time triplet-state response.

Because the singlet-state nonadiabatic dynamics did not explicitly include spin--orbit coupling or intersystem crossing, we generated a representative $T_1$ nuclear ensemble by propagating independent trajectories on the $T_1$ potential-energy surface (\cref{sec:ComputationalDetails}).
The vibrationally averaged MR-ADC(2) spectrum (\cref{fig:T1_vib_avg}) retains the principal features of the corresponding single-point spectrum while broadening the individual bands and redistributing their intensities.
Both calculations reproduce peak 1 near 281.4 eV, which is assigned predominantly to a C$_3$ 1s excitation into the singly occupied $\pi$-type orbital.
Vibrational averaging shifts this feature slightly to higher energy and broadens it relative to both experimental spectra.
More substantial discrepancies appear at higher energies: MR-ADC(2) markedly underestimates the intensity of peak 2 near 283.8 eV and overestimates the magnitude of the ground-state bleach.
The persistence of these deviations relative to both experimental datasets points to limitations in the electronic structure description rather than to differences between the measurements.

The $T_1$ spectrum is considerably more sensitive to the MR-ADC approximation than the ground-state spectrum.
Whereas MR-ADC(2) and MR-ADC(2)-SX produce nearly identical ground-state XAS spectra, inclusion of first-order semi-internal couplings in MR-ADC(2)-SX leads to substantial changes in both the energies and intensities of the triplet-state features (\cref{fig:nto_triplet}).
Most notably, MR-ADC(2)-SX recovers much of the missing intensity of peak 2 and yields a substantially improved balance between peaks 1 and 2 in comparison with both experiments.
Natural transition orbital analysis assigns peak 1 primarily to a C$_3$ 1s excitation into a singly occupied $\pi$-type orbital delocalized over the central conjugated framework, whereas peak 2 is dominated by a C$_4$ 1s excitation into a related low-lying $\pi$-type orbital.
The higher-energy region contains additional contributions from carbonyl-centered C~1s excitations into higher-lying $\pi^\ast$ orbitals.
MR-ADC(2)-SX also provides a noticeably improved description of the structure near peak 4 and of the ground-state bleach when compared with the higher-resolution Haugen spectrum.
These differences indicate that the open-shell core-excited states contributing to the $T_1$ spectrum are particularly sensitive to correlation and orbital-relaxation effects.

The resulting assignments are broadly consistent with the TD-DFT interpretation of Bhattacherjee et al.\cite{bhattacherjee:2017p16576} and the coupled-cluster analysis of Faber et al.,\cite{faber:2019p144107} which likewise associate the long-time spectral features with carbon 1s excitations into triplet-state $\pi$ orbitals.
As in the coupled-cluster calculations, MR-ADC(2) underestimates the intensity of the second low-energy triplet feature, while MR-ADC(2)-SX recovers much of this missing intensity and improves the overall spectral profile.
The agreement with both long-time experimental spectra reinforces the assignment of peaks 1 and 2 as distinct, carbon-site-specific signatures of the $T_1$ state.
We further note that the MR-ADC(2)-SX spectrum agrees well with the single-point TR-XAS spectrum reported by Tsuru et al.\cite{tsuru:2021p024101} using a low-spin open-shell reference EOM-CCSD approach, providing additional support for the revised triplet-state description.

\section{Conclusions}
\label{sec:Conclusion}

Time-resolved X-ray absorption spectroscopy (TR-XAS) offers an element- and site-specific probe of coupled electronic and nuclear dynamics, but its interpretation requires methods that can treat multiconfigurational excited states across nonequilibrium nuclear ensembles.
Here, we combined surface-hopping molecular dynamics with multireference algebraic diagrammatic construction (MR-ADC), a Green's-function-based excited-state method built on a multiconfigurational reference and a systematic treatment of electron correlation.
This framework enabled trajectory-averaged simulations of the carbon K-edge spectrum of acetylacetone (AcAc) throughout its relaxation following $S_2$ excitation (0 -- 200 fs).
Benchmark calculations identified MR-ADC(2)/CASSCF(10e,8o)/cc-pwCVDZ as an accurate and computationally tractable protocol.
For the ground-state XAS spectrum, nuclear averaging has only a modest effect below 287 eV but substantially improves the higher-energy spectrum, where transitions involving the terminal methyl groups are more sensitive to molecular motion.

The early- (20--100 fs) and intermediate-time (120--200 fs) spectra demonstrate that the TR-XAS response cannot be interpreted solely from electronic-state populations or single-point calculations at selected stationary geometries.
Instead, the observed spectral evolution reflects heterogeneous $S_2$ and $S_1$ ensembles undergoing excited-state intramolecular hydrogen transfer, bond alternation, and ring opening.
The weak feature at 279.5--281.5 eV emerges as a spectroscopic marker of short-lived, approximately $C_{2v}$-symmetric proton-sharing structures, in which localization of the accepting $\pi$-type orbital on C$_3$ enhances the corresponding C$_3$ 1s transition.
The prominent 284--286 eV doublet likewise cannot be assigned uniquely to Franck--Condon and relaxed excited-state structures.
Our calculations instead suggest a broad, geometry-dependent manifold of excited-state transitions, whose overlap with the ground-state bleach contributes to the apparent separation into two maxima.

At longer delays, the calculated $T_1$ spectra reproduce the principal experimental triplet-state features and assign the absorptions near 281.4 and 283.8 eV primarily to C$_3$ and C$_4$ 1s excitations into low-lying triplet-state $\pi$ orbitals.
The triplet spectrum is considerably more sensitive to the MR-ADC approximation than the ground-state spectrum: MR-ADC(2)-SX restores much of the intensity underestimated by MR-ADC(2) near 283.8 eV and provides a more balanced description of the first two experimental peaks.
Overall, these results demonstrate that explicit sampling of coupled electronic and nuclear dynamics, combined with a multireference treatment of core-excited states, is essential for extracting structural and mechanistic information from the time-dependent carbon K-edge spectrum of AcAc.

\section*{Conflicts of Interest}
There are no conflicts to declare.

\section*{Data Availability}
The data that supports the findings of this study are available within the article and its supplementary material. 
Additional data can be made available upon reasonable request.

\section*{Acknowledgements}
This material is based upon work supported by the U.S. Department of Energy, Office of Science, Office of Basic Energy Sciences, Chemical Sciences, Geosciences, and Biosciences Division, Atomic, Molecular, and Optical Sciences Program, under Award Number DE-SC0026341.
Additionally, B.W.C. was supported through the NSF REU grant No.~CHE-2150102.
Computations were performed at the Ohio Supercomputer Center under Project No.\@ PAS1583.
\cite{OhioSupercomputerCenter1987}

%%%END OF MAIN TEXT%%%

%The \balance command can be used to balance the columns on the final page if desired. It should be placed anywhere within the first column of the last page.

\balance

%If notes are included in your references you can change the title from 'References' to 'Notes and references' using the following command:
%\renewcommand\refname{Notes and references}

%%%REFERENCES%%%
%\bibliography{acac_paper}
%\bibliographystyle{rsc}

\providecommand*{\mcitethebibliography}{\thebibliography}
\csname @ifundefined\endcsname{endmcitethebibliography}
{\let\endmcitethebibliography\endthebibliography}{}

\end{document}